\documentclass[fleqn,usenatbib]{mnras}

\usepackage{newtxtext,newtxmath}

\usepackage[T1]{fontenc}
\usepackage{lscape}
\usepackage{rotating}
\usepackage{siunitx}
\usepackage{adjustbox}

\usepackage{graphicx}	
\usepackage{amsmath}	
\usepackage{amssymb}	

\title[Dust Emissivity Index in M31]{Investigating variations in the dust emissivity index in the Andromeda galaxy}

\author[G. Athikkat-Eknath]{
G. Athikkat-Eknath,$^{1}$\thanks{E-mail: g.athikkat.eknath@gmail.com}
S. A. Eales,$^{1}$
M. W. L. Smith,$^{1}$
A. Schruba,$^{2}$
K. A. Marsh$^{3}$
and
A. P. Whitworth$^{1}$
\\
$^{1}$School of Physics and Astronomy, Cardiff University,
          Queen's Buildings, The Parade, Cardiff CF24 3AA, UK\\
$^{2}$Max-Planck-Institut f\"{u}r Extraterrestrische Physik, Giessenbachstr. 1, 85748 Garching, Germany\\
$^{3}$Infrared Processing and Analysis Center, CalTech, 1200E California Boulevard Pasadena, CA91125, USA
}

\date{Accepted 2021 October 22. Received 2021 October 22; in original form 2021 March 26}

\pubyear{2021}

\begin{document}
\label{firstpage}
\pagerange{\pageref{firstpage}--14}
\maketitle

\begin{abstract}

  Over the past decade, studies of dust in the Andromeda galaxy (M31) have shown radial variations in the dust emissivity index ($\beta$). Understanding the astrophysical reasons behind these radial variations may give clues about the chemical composition of dust grains, their physical structure, and the evolution of dust. We use $^{12}$CO(J=1-0) observations taken by the Combined Array for Research in Millimeter Astronomy (CARMA) and dust maps derived from \textit{Herschel} images, both with an angular resolution of 8" and spatial resolution of 30 pc, to study variations in $\beta$ across an area of $\approx$ 18.6 kpc$^2$ in M31. We extract sources, which we identify as molecular clouds, by applying the \textsc{astrodendro} algorithm to the $^{12}$CO and dust maps, which as a byproduct allows us to compare continuum emission from dust and CO emission as alternative ways of finding
  molecular clouds. We then use these catalogues to
  investigate whether there is evidence that $\beta$ is different inside and outside
  molecular clouds. Our results confirm the radial variations of $\beta$ seen
  in previous studies.
  However, we find little difference between the average $\beta$ inside molecular clouds compared to outside molecular clouds, in disagreement with models which predict an increase of $\beta$ in dense environments. Finally, we find some clouds traced by dust with very little CO which may be either clouds dominated by atomic gas or clouds of molecular gas that contain little CO.

\end{abstract}




\section{Introduction}
\label{sec:intro}

Although primarily known for its obscuration effects on starlight, interstellar dust plays many other important roles in the interstellar medium (ISM): heating the ISM through the photoelectric effect (\citealt{Draine1978}), providing a cooling mechanism via its infrared emission in dense regions, shielding molecules from dissociating radiation, and even providing a surface for catalysing the formation of hydrogen molecules (\citealt{Gould1963}, \citealt{Hollenbach1971}). Many authors have suggested that the continuum emission from interstellar dust can be used to trace the mass of gas in galaxies (e.g. \citealt{Eales2012}, \citealt{Magdis2012}, \citealt{Liang2018}, \citealt{Groves2015}, \citealt{Scoville2016}, \citealt{Scoville2017}, \citealt{Tacconi2018}, \citealt{Janowiecki2018}), as an alternative to traditional tracers like carbon monoxide (CO). However, this method can only work if the properties of dust are universal, or if their variation with environment or epoch is known.

The dust emissivity index ($\beta$) is an important property that acts as a modifier to the shape of the blackbody spectrum which describes the emission from dust. In the optically thin limit (where the optical depth $\tau \ll$ 1), the specific intensity of dust emission is given by:
\begin{equation}
\label{eq:1}
  I_{\nu} \propto B_{\nu}(T)\nu^{\beta}
\end{equation}
where $B_{\nu}(T)$ is the Planck function and $\nu$ is the frequency. \cite{Smith2012} have used {\it Herschel} observations to investigate the properties of the dust
in the Andromeda galaxy (M31),
discovering that $\beta$ varies radially within the galaxy's disk. This general trend has been confirmed by \cite{Draine2014} using independent \textit{Herschel} data for M31 but with a different method. Variations in $\beta$ have also been reported in two other large spiral galaxies within the Local Group, the Milky Way (MW) and the Triangulum galaxy (M33). In M33, there is evidence for radial variations in $\beta$ and dust temperature, both of which decrease with galactocentric radius (\citealt{Tabatabaei2014}). So far, however, we have very little understanding of what is causing such radial variations. \cite{Tabatabaei2014} have found that $\beta$ is higher in regions where there is molecular gas traced by $^{12}$CO(J=2-1) or strong H$\alpha$ emission in M33 but the authors did not investigate whether there is any difference in $\beta$ between low-density and high-density environments at
the same radius. In the MW, the \textit{Planck} team have shown that $\beta$ decreases from smaller Galactic longitudes to larger Galactic longitudes (\cite{PlanckCollaboration2014b}; see their Fig. 9). They have also found
an increase in $\beta$ by $\sim$0.23 in the regions dominated by molecular gas along the line of sight when compared to the more diffuse atomic medium (\cite{PlanckCollaboration2014a}; see their Fig. 12); although as in M33, it is
not clear whether the increase of $\beta$ in dense environments is the explanation of any radial variation in $\beta$.

Radial variations in $\beta$ have also been seen in galaxies outside the Local Group. In a sample of 61 galaxies from Key Insights into Nearby Galaxies: Far-Infrared Survey with \textit{Herschel} (KINGFISH), \cite{Hunt2015} have found that the radial effects of $\beta$ can vary from one galaxy to another (see their Fig. 10). For example, some galaxies show negative radial gradients (e.g. NGC0337, NGC3049, NGC3077, NGC4559, NGC4725) whereas others show positive radial gradients (e.g. NGC1482, NGC3773, NGC4321, NGC4594). Evidence for  variation in the global values
of $\beta$ has come from a study of the gas and dust in 192 galaxies
in the JCMT dust and gas In Nearby Galaxies Legacy Exploration (JINGLE) project (\citealt{Lamperti2019}, \citealt{Smith2019}). The authors find  correlations between $\beta$ and properties such as stellar mass, stellar mass surface density, metallicity, H{\sc i} mass fraction, star formation rate (SFR), specific SFR, SFR surface density, and the ratio of SFR and dust mass for these galaxies. The strongest positive correlation is found between $\beta$ and stellar mass surface density.

Our study focuses on variations in dust properties, in particular the dust emissivity index ($\beta$), within M31. Due to its proximity to us \citep[at a distance of $\approx$ 785 kpc;][]{McConnachie2005}, we can learn about properties of dust and the ISM in M31 at the scale of individual molecular clouds. M31 also provides a unique perspective as the biggest spiral galaxy in the Local Group, with the added incentive that we can observe the galaxy from the outside, unlike observing the Milky Way from within which limits us from getting a global view of our Galaxy and has problems such as superimposed sources at different distances along the line-of-sight.

There are many archival datasets containing observations of M31 in different wavebands (\citealt{Thilker2005}, \citealt{Braun2009}, \citealt{Dalcanton2012}, \citealt{PlanckCollaboration2015}), including key far-infrared and sub-mm observations from the Herschel Exploitation of Local Galaxy Andromeda (HELGA) survey (\citealt{Smith2012}, \citealt{Fritz2012}) and $^{12}$CO(J=1-0) observations (\citealt{Nieten2006}) made using the 30-m telescope at the Institut de Radioastronomie Millim\'etrique (IRAM) to trace the distribution of molecular gas over the whole galaxy. We take advantage of the high resolution observations covering part of M31 obtained with the Combined Array for Research in Millimeter-wave Astronomy (CARMA) inteferometer to trace molecular gas (A.~Schruba et al, in preparation). The central research question this work attempts to address is: \textbf{`Are the radial variations in the dust emissivity index ($\beta$) in the Andromeda galaxy caused by an increase of $\beta$ in dense molecular gas regions?'} As such, we measure and compare $\beta$ in dense molecular gas regions with $\beta$ in non-dense regions.

Given the evidence from {\it Fermi} (\citealt{Abdo2010}), {\it Planck} (\citealt{PlanckCollaboration2011}) and {\it Herschel} (\citealt{Pineda2013}) that there is CO-dark molecular gas in the Milky Way, probably because of photodissociation of the CO molecule \citep{Hollenbach1997}, we trace clouds in two different ways, using CO emission and dust continuum emission. This allows us to carry out the additional interesting project of comparing
the cloud catalogues produced by the two different methods.

This paper is structured as follows: we first describe the observational data that are used in this work (Section \ref{sec:obs}). Next, we outline our source extraction methodology (Section \ref{sec:se}), followed by our results (Section \ref{sec:res}) and discussion (Section \ref{sec:disc}). Finally, we summarise our conclusions in Section \ref{sec:conc}.

\section{Observations}
\label{sec:obs}
\subsection{CARMA survey of Andromeda}
\begin{figure*}
 \centering
 \includegraphics[width=17cm]{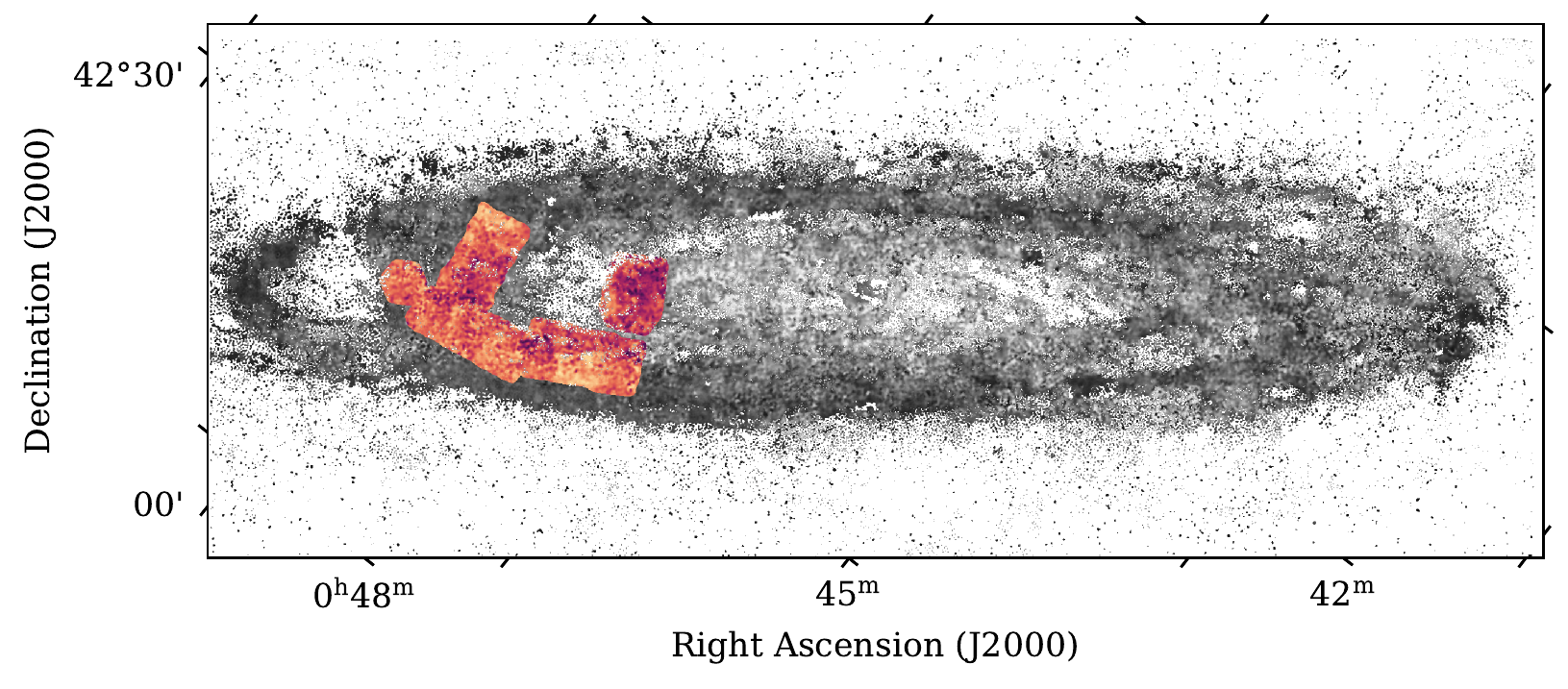} \caption{Map of dust emissivity index ($\beta$) made from PPMAP. The maroon-yellow part shows the region of the $\beta$ map overlapping the $^{12}$CO(J=1-0) map made with CARMA, partially covering the inner ring at 5 kpc and the dusty, star-forming ring at 10 kpc.}
 \label{fig:carmareg}
 \end{figure*}

We use a map of $^{12}$CO(J=1-0) integrated intensity obtained from observations of M31 made using the CARMA interferometer. The data were taken as part of the `CARMA Survey of Andromeda' (A.~Schruba et al., in preparation) across an angular area of 365 arcmin$^2$ and on-sky physical area covering 18.6 kpc$^2$, which includes parts of M31's inner 5 kpc gas ring and 10 kpc dusty, star-forming ring (\citealt{Habing1984}). This corresponds to a deprojected physical area of $\approx$ 84.6 kpc$^2$ at our chosen distance to M31 (785 kpc) and inclination (77$^{\circ}$). The region covered in our work is highlighted in Figure \ref{fig:carmareg}. The high-resolution CARMA data were combined with observations from the IRAM 30 m telescope (\citealt{Nieten2006}), to capture the emission at large angular scales missed by the interferometric
data (\citealt{Caldu-Primo2016}, A.~Schruba et al. in preparation). Without this correction, 43\% of the CO flux would have been lost \citep{Caldu-Primo2016}. The data were merged using the \texttt{immerge} task from the data reduction software \textsc{miriad} (\citealt{Sault1995}), which performs a linear combination of the low resolution and high resolution data cubes in Fourier space. The data
were merged using unit weights for the single-dish data at all spatial frequencies, which leaves the CARMA beam unchanged (A.~Schruba et al. in preparation).
 The merged CARMA + IRAM data have a pixel scale size of 2" and the beam width is approximately 5.5" or 20 pc. For our analysis, we convolve the 5.5" $^{12}$CO(J=1-0) map with a Gaussian kernel of full width half maximum (FWHM) $\theta_\mathrm{kernel}$ = $\sqrt{\theta^{2}_{8"} - \theta^{2}_{5.5"}}$ = 5.8" to obtain a resulting map with a FWHM of 8" - the effective resolution of our dust observations (see Section \ref{ssec:ppmapobs}). The map is then reprojected to match the 4" pixel scale size of our dust maps in order to ensure consistent pixel-by-pixel analysis across the CO and dust data.

\cite{Caldu-Primo2016} give a 1$\sigma$ error in the molecular gas mass surface density derived from their CO observations, on the assumption of a line width of
10 km s$^{-1}$, of 0.83 M$_{\odot}$ pc$^{-2}$. We have estimated the error in
the molecular column density using the robust empirical technique of calculating the
standard deviation in groups of pixels. This is a conservative technique since the
error will also include a contribution from the variance in the distribution of
the molecular gas. We find no evidence for a variation in sensitivity across the
image and estimate that the 1$\sigma$ sensitivity of our CO integrated intensity map is 1.20 $\mathrm{K \; km \; s}^{-1}$. If we adopt a constant conversion factor, $X_{\mathrm{CO}}$ = 1.9 $\times$ 10$^{20}$ cm$^{-2}$ [K km s$^{-1}$]$^{-1}$ (\citealt{Strong1996}), this corresponds to a molecular gas mass surface density of 3.63 M$_{\odot}$pc$^{-2}$. We do not account for helium in our molecular gas mass surface density ($\Sigma_{\mathrm{H_2}}$) calculations. The molecular gas mass surface density including helium can be calculated using $\Sigma_{\mathrm{gas, mol}} = 1.36 \times \Sigma_{\mathrm{H_2}}$.

\subsection{PPMAP algorithm applied to Herschel observations}
\label{ssec:ppmapobs}
\cite{Smith2012} have produced dust temperature ($T_{\mathrm{dust}}$), dust emissivity index ($\beta$) and dust mass surface density ($\Sigma_{\mathrm{dust}}$) maps of M31 by applying spectral energy distribution (SED) fitting to continuum emission observations made by \textit{Herschel} centred at wavelengths of 70, 100, 160 and 250, 350, 500 $\mu$m. In their work, \cite{Smith2012} convolved all \textit{Herschel} images to the resolution of the lowest resolution image and made the assumption that all dust was at a single temperature along the line of sight.

Since then, \cite{Marsh2015} have created a way of increasing the spatial resolution of maps of the dust properties through a procedure called point process mapping (PPMAP). When applied to the \textit{Herschel} data, PPMAP provides data products with an angular resolution of 8",  roughly corresponding to the resolution of the shortest \textit{Herschel} wavelength (70 $\mu$m). By removing the requirement
to convolve the images to \textit{Herschel}'s lowest angular resolution of 36" (at 500 $\mu$m), PPMAP allows us to probe spatial scales of $\approx$ 30 pc.

PPMAP discards the assumption that the $T_{\mathrm{dust}}$ and $\beta$ along the line of sight is uniform, with the only key assumption being that dust emission is in local thermal equilibrium and is optically thin. Through the application of PPMAP to \textit{Herschel} observations, \cite{Whitworth2019} have produced $T_{\mathrm{dust}}$, $\beta$ and $\Sigma_{\mathrm{dust}}$ maps of M31 (see their Fig. 2 (b) and (c)) which we use in our analysis in this paper. PPMAP produces estimates of the dust mass surface density in each pixel at a set of discrete dust temperatures and emissivity values. In creating their dust maps, \cite{Whitworth2019} have used twelve values of dust temperature ($T_{\mathrm{dust}}$ = 10.0 K, 11.6 K, 13.4 K, 15.5 K, 18.0 K, 20.8 K, 24.1 K, 27.8 K, 32.2 K, 37.3 K, 43.2 K, 50.0 K) and four values of dust emissivity index ($\beta$ = 1.5, 2.0, 2.5 and 3.0). We have averaged over these intervals to create a map of $\Sigma_{\mathrm{dust}}$ and maps of the mass-weighted $T_{\mathrm{dust}}$ and $\beta$ values.
We choose to perform our analysis on the average $T_{\mathrm{dust}}$, $\beta$ and $\Sigma_{\mathrm{dust}}$ maps because these
are what we need for comparison with the work of \cite{Smith2012} and \cite{Draine2014}. This is also a more conservative approach than using the data in each PPMAP slice as, although \cite{Marsh2015} have
tested the PPMAP slice results using synthetic observations of the Milky Way, there have been no such tests done using synthetic observations on extragalactic scales.
 Further details about the PPMAP algorithm can be found in the work of \cite{Marsh2015}, with the full method for creating the M31 maps described in \cite{Whitworth2019}. We have removed a scaling factor from the $\Sigma_{\mathrm{dust}}$ map for our analysis\footnote{PPMAP produces a map of the total mass surface density of interstellar matter (gas + dust) using a dust mass opacity coefficient of $\kappa_{300, \; \mathrm{PPMAP}} = 0.010 \; \mathrm{m^2 \; kg^{-1}}$. Here we convert the PPMAP results to the surface density of dust alone using $\kappa_{350}$ = 0.192 m$^2$ kg$^{-1}$ (\citealt{Draine2003}) and a gas-to-dust ratio of 100.}.

The minimum
value in our map of dust mass surface density is $\Sigma_{\mathrm{dust}} \simeq 0.05$  M$_{\odot}$ pc$^{-2}$, which corresponds roughly to a 5$\sigma$ detection.
We focus on the CARMA-observed region within the PPMAP dust maps (see Figure \ref{fig:carmareg}) for ease of comparison of dust properties with observations of $^{12}$CO(J=1-0). The rise in $\beta$ with galactocentric radius in the central 3 kpc of M31 followed by a fall in $\beta$ beyond this radius, first seen by \cite{Smith2012}, has been confirmed by \cite{Whitworth2019} in their reanalysis of the original \textit{Herschel} data using PPMAP.

\subsection{H{\sc i} ancillary data}
\label{ssec:HIobs}
We use the H{\sc i}  column density map of M31 obtained by \cite{Braun2009} using the Westerbork Synthesis Radio Telescope (WSRT) array at an angular resolution of 30" and spatial resolution of $\approx$ 110 pc. The observed H{\sc i} in M31 has a smooth distribution with a `clumping factor' of $\approx$ 1.3  (\citealt{Leroy2013}). Therefore, although these data do not match the resolution of our CO and dust maps, the H{\sc i} map provides an estimate of the contribution of
the smooth HI distribution at the position of a molecular cloud. We emphasise that our estimate of the H{\sc i} contribution at the scale of an individual cloud is at best a rough estimate. The H{\sc i} map has not been corrected for opacity effects as the best method of doing this is still uncertain. Localised opacity corrections can lead to an increase of the inferred H{\sc i} gas mass by $\approx$ 30\% or more (e.g. \citealt{Braun2009}, \citealt{Koch2021}), and so this an additional uncertainty in our estimates of the contribution of HI at the position of a cloud. The data have a pixel scale size of 10" which we reproject into 4" $\times$ 4" pixels to match the PPMAP map projection. We focus on the CARMA-observed region within this reprojected map. The 1$\sigma$ sensitivity of the H{\sc i} column density map is 4.05 $\times$ 10$^{19}$ cm$^{-2}$, corresponding to an atomic gas mass surface density value of 0.32 M$_{\odot}$ pc$^{-2}$. We do not account for helium in our atomic gas mass surface density ($\Sigma_{\mathrm{H{\textsc i}}}$) calculations.

\subsection{Astrometric offsets}
We estimate the astrometric accuracy of the instruments as FWHM of the beam divided by the signal-to-noise ratio of the pointing source observations. Typical pointing calibrator observations reach a signal-to-noise level of $\approx$ 10. Therefore, we estimate that the CARMA object positions are accurate to within 0.5”. The IRAM astrometric offset should not be more than 2.3”. We expect that the positions of objects from the raw \textit{Herschel} maps are accurate to 2”\footnote{SPIRE astrometry-corrected maps readme: \url{http://archives.esac.esa.int/hsa/legacy/HPDP/SPIRE/SPIRE-P/ASTROMETRY/README.html}}. We do not think these astrometric offsets are likely to be significant because the smallest area of a cloud in our source extraction has been taken as 10 pixels (see Table \ref{tab:param}), making a cloud much larger than the beam of either the CO map and the effective beam of the dust map.

\section{Source extraction}
\label{sec:se}
\subsection{Dendrogram}

\begin{table*}
\caption{Input dendrogram parameters for source extraction.}              
\label{tab:param}      
\centering                                      
\begin{tabular}{c c c c}          
\hline\hline                        
Data & Minimum value  & Minimum structure significance value & Minimum no. of pixels \\    
& \texttt{min\_value} & \texttt{min\_delta} & \texttt{min\_npix} \\
\hline                                   
     CO & 4.57 K\,km s$^{-1}$ & 2.40 K\,km s$^{-1}$ & 10 \\      
     Dust & 0.44 M$_{\odot}$ pc$^{-2}$ & 0.296 M$_{\odot}$ pc$^{-2}$ & 10 \\
\hline                                             
\end{tabular}
\end{table*}

A common method for accessing the hierarchical structure of molecular clouds is to use a dendrogram. A dendrogram allows us to segregate the denser regions from the more diffuse regions and access any nested sub-structure. We compute dendrograms for the CO and PPMAP maps using the Python package \textsc{astrodendro 0.2.0} (\citealt{Rosolowsky2008}). This package allows us to construct an empirically motivated segmentation of `clouds' within our data.
If we adopt the analogy within the \textsc{astrodendro} documentation\footnote{\textsc{astrodendro} documentation: \\ \url{https://dendrograms.readthedocs.io/en/stable/}}, a dendrogram can be represented as a "tree" with a trunk, and the nested structures of this trunk are called "branches" which contain sub-structures in the form of "leaves".  The resulting sources (molecular clouds) extracted by our computed dendrograms are analogous to leaves on the tree.

The dendrogram computation requires the specification of three parameters:
\begin{enumerate}
  \item The minimum intensity value of a pixel (\texttt{min\_value}): the dendrogram will discard any pixels fainter than this threshold.
  \item The minimum significance value for a leaf (nested structure) to be identified as an independent object (\texttt{min\_delta}): if the difference between a new local maximum pixel value (peak of prospective structure) and the last pixel value examined in an existing structure (point at which the new structure may be merged onto the existing one) is greater than this parameter, a structure is considered to be significant enough to be independent.
  \item The minimum size (defined in number of pixels) required to identify a structure as an independent object (\texttt{min\_npix}): if the number of pixels in a structure does not match or exceed this value, the structure is merged with an existing structure.
\end{enumerate}

For our work, we select these parameter values carefully as described in Sections \ref{ssec:gmc_co} and \ref{ssec:gmc_ppmap}. Further details about the dendrogram algorithm can be found in the documentation of the \textsc{astrodendro} package and in the work of \cite{Rosolowsky2008}.

\subsection{Identifying molecular clouds with CO}
\label{ssec:gmc_co}
We choose to run the dendrogram on the CO integrated intensity map rather than trying to find clouds in the original CO data cube. Our reasoning for this is as follows:
firstly, although the CO linewidth information would allow us to distinguish whether a cloud found by the dendrogram is a single source or multiple clouds along the line-of-sight, knowledge of the 3D structure of the gas does not provide any benefit in understanding the cause of the spatial variation in dust emissivity index as we do not have access to matching information on our dust emissivity index map. Moreover, the CO method could still suffer from the problem that approximately one-third of the molecular gas may not contain any CO (see Section \ref{sec:intro}) and so would not give us a perfect benchmark catalogue of clouds to compare all other catalogues with. In order to make as direct a comparison as possible of the two tracers (CO and dust) for finding clouds, we compute the dendrogram on the CO integrated intensity map.

The input parameters used for our source extraction are listed in Table \ref{tab:param}. To calculate the 1$\sigma$ noise threshold, we select a region within our smoothed and reprojected map where we see no obvious sources and calculate the standard deviation ($\sigma_{\mathrm{std}}$) of pixel values. Our minimum detection threshold (\texttt{min\_value}) is 3$\sigma$, which we add to a mean background level of 0.97 $\mathrm{K \; km \; s}^{-1}$ to account for a diffuse constant level of CO present beneath denser structures. For our minimum structure significance threshold (\texttt{min\_delta}), we choose a value of 2$\sigma$. This means that clouds will
 have a peak CO intensity of at least 5$\sigma$ above the mean background level. The final key requirement for our dendrogram extraction is that an independent structure should have a minimum size (\texttt{min\_npix}) of 10 pixels. This size threshold is larger than the beam area (calculated using the diameter as the FWHM = 8") divided by the area of one pixel.

 Clouds are defined as the objects at the highest level of the dendrogram hierarchy (i.e. dendrogram `leaves'), containing no nested sub-structures. We find 140 sources from our CO observations (see Figure \ref{fig:carma_cons_sources}) which we assume are molecular clouds traced by CO.

\subsubsection{Determining the gas and dust properties of clouds}
\label{sssec:gdpropmethod}
The molecular gas mass surface density in a pixel, $\Sigma_{\mathrm{H_2}}$, is related to the CO intensity in the pixel, $I_{\mathrm{CO}}$, by:
\begin{equation}
\label{eq:2}
\Sigma_\mathrm{H_2} = I_\mathrm{CO} \times X_\mathrm{CO} \times m(\mathrm{H_2})
\end{equation}
where $m(\mathrm{H_2})$ is the mass of a hydrogen molecule. We adopt a constant $X_{\mathrm{CO}}$ = 1.9 $\times$ 10$^{20}$ cm$^{-2}$ [K km s$^{-1}$]$^{-1}$ from \cite{Strong1996}. We calculate the total CO-traced molecular gas mass in each cloud by multiplying the gas mass surface density values by the area of a pixel and summing over all gas mass values in the dendrogram leaf.

We also calculate the dust-mass-weighted mean $T_{\mathrm{dust}}$ and mean $\beta$ for our clouds using the pixels corresponding to our dendrogram leaves in these maps.
We calculate the total dust mass in a cloud by multiplying the $\Sigma_{\mathrm{dust}}$ values from the PPMAP map by the area of a pixel and summing over all dust mass values in the dendrogram leaf. The CO-traced molecular gas-to-dust ratio (GDR) is obtained for each cloud by:
\begin{equation}
  \label{eq:3}
    \mathrm{ CO\mbox{-}traced \; molecular \; GDR} = \frac{M_{\mathrm{H_2}}}{M_{\mathrm{dust}}}
\end{equation}
where $M_{\mathrm{H_2}}$ is the total CO-traced molecular gas mass in the cloud, and $M_{\mathrm{dust}}$ is the total mass of dust in the cloud.

The gas mass surface density of atomic hydrogen in a pixel is given by:
\begin{equation}
\label{eq:4}
  \Sigma_{\mathrm{H{\textsc i}}} = N(\mathrm{H}) \times m(\mathrm{H})
\end{equation}
where $N$(H) is the column density of atomic hydrogen obtained from the H{\textsc i} map and $m(\mathrm{H})$ is the mass of a hydrogen atom. The total atomic gas mass in the cloud, $M_{\mathrm{HI}}$, is calculated by multiplying $\Sigma_{\mathrm{H{\textsc i}}}$ with the area of a pixel and summing over all atomic gas mass values in the dendrogram leaf. The total (molecular + atomic) GDR is obtained by:
\begin{equation}
  \label{eq:5}
  \mathrm{ Total \; GDR} = \frac{(M_{\mathrm{H{\textsc i}}} + M_{\mathrm{H_2}})}{M_{\mathrm{dust}}}
\end{equation}

To propagate the error in molecular GDR values, we make the approximation that the noise from the CO map dominates over error contributions from PPMAP dust measurements. To propagate the error in total GDR values, we add the noise from both the CO and H{\sc i} maps in quadrature. We neglect systematic errors from PPMAP. The error in the CO intensity and H{\sc i} column density within each cloud are calculated using $\mathrm{N}_{\mathrm{pix, beam}} \times  \sqrt{\mathrm{N}_{\mathrm{pix, cloud}}/\mathrm{N}_{\mathrm{pix, beam}}} \times \sigma_{\mathrm{std}}$, where $\sigma_{\mathrm{std}}$ is the standard deviation of pixel values within a region of each map where we see no obvious sources. $\mathrm{N}_{\mathrm{pix, beam}}$ is number of pixels in the beam and $\mathrm{N}_{\mathrm{pix, cloud}}$ is the number of pixels in a cloud.

We calculate the radial distance of a cloud from the centre of M31 (RA: 00$^h$ 42$^m$ 44.33$^s$, Dec: 41$^{\circ}$ 16' 7.5" (\citealt{Skrutskie2006}))
assuming that the pixel with the peak intensity marks the centre of the cloud. The distance to Andromeda used in this work is 785 kpc (\citealt{McConnachie2005}) and Andromeda's inclination angle is 77$^{\circ}$ (\citealt{Fritz2012}).

\subsubsection{Determining dust properties in non-dense regions}
We compare the values of $\beta$ and $T_{\mathrm{dust}}$ in the pixels within our clouds with the values of the pixels that fall outside the clouds (non-dense regions). We identify all the pixels which fall within the non-dense regions by masking out pixels within our dendrogram leaves. We split the pixels into two radial bins (5-7.5 kpc and 9-15 kpc) and create histograms of $T_{\mathrm{dust}}$ and $\beta$ values at these radii, both inside and outside dense regions (see Section \ref{ssec:radvarres}).

\subsection{Identifying molecular clouds with dust}
\label{ssec:gmc_ppmap}

\begin{figure}
\hspace{-2cm}
\includegraphics[width=12.25cm]{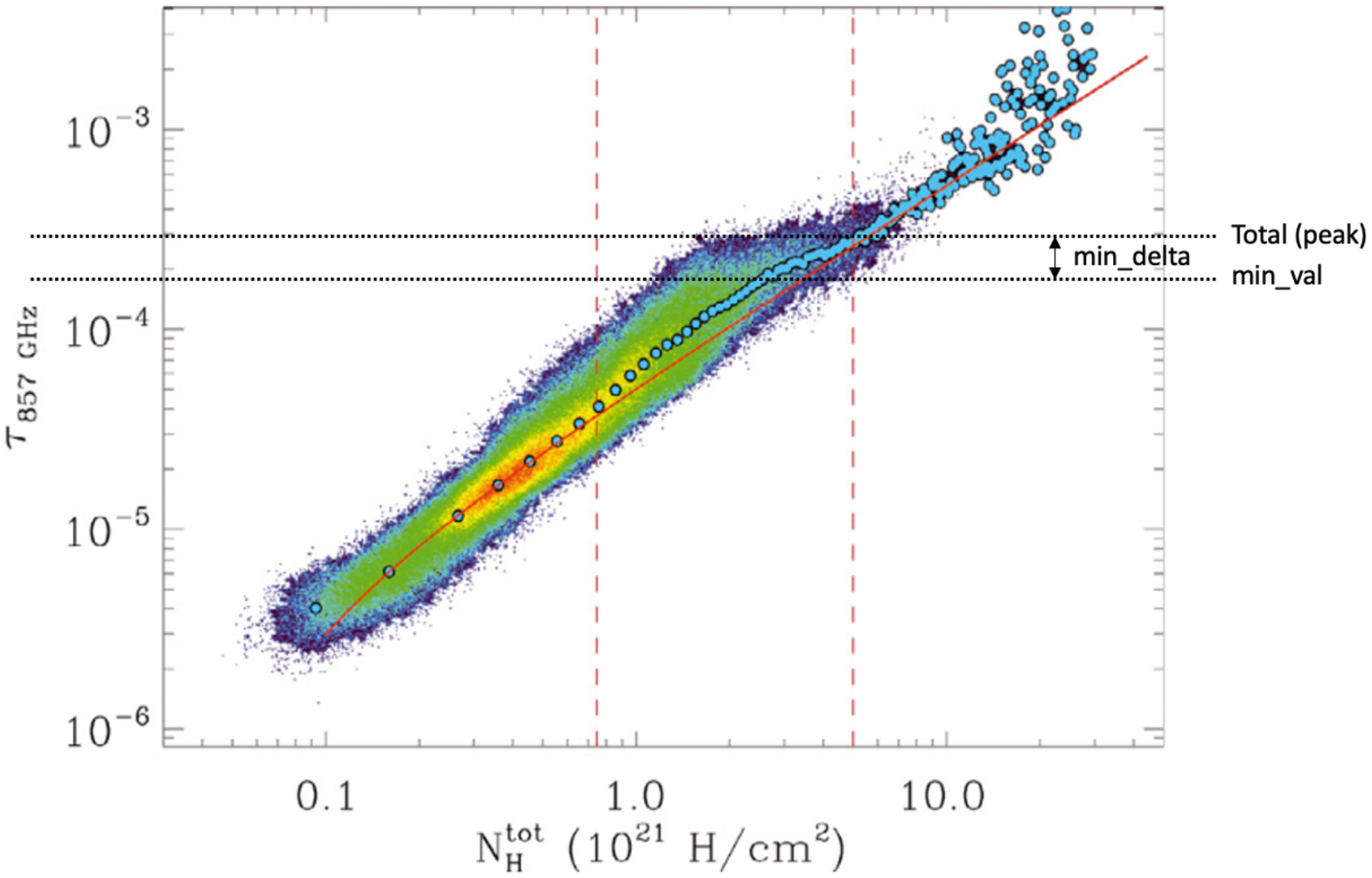}
\vspace{-1.75cm}
\caption{Adaptation of Fig. 6 from \citet{PlanckCollaboration2011} showing the correlation between the dust optical depth $\tau$ at 857 GHz and the total gas column density. The dashed vertical red line at $\mathrm{N}_{\mathrm{H}}^{\mathrm{tot}} \approx 8.0 \times 10^{20}$ H cm$^{-2}$ shows the threshold above which excess thermal emission from dust traces CO-dark molecular gas. The dashed vertical red line at N$_{\mathrm{H}}^{\mathrm{tot}} \approx 5.0 \times 10^{21}$ H cm$^{-2}$ shows the threshold at which the gas column density becomes dominated by the molecular gas traced by CO emission. The lower black dotted horizontal line shows the $\tau_{857 \; \mathrm{GHz}}$ value used to calculate the \texttt{min\_value} dendrogram parameter for our $\Sigma_{\mathrm{dust}}$ map. The difference between the black dotted lines shows the $\tau_{857 \; \mathrm{GHz}}$ value used to calculate the \texttt{min\_delta} dendrogram parameter. All of our clouds have a peak dust mass surface density of $\Sigma_{\mathrm{dust}} \geq 0.74 \; \mathrm{M}_{\odot} \mathrm{pc}^{-2}$ (Equation \ref{eq:6}), equivalent to a peak optical depth of $\tau_{857 \; \mathrm{GHz}} \geq 3.0 \; \times 10^{-4}$ (upper black dotted line).}
\label{fig:planck_cartoon}
\end{figure}

\begin{figure*}
\includegraphics[width=16.5cm]{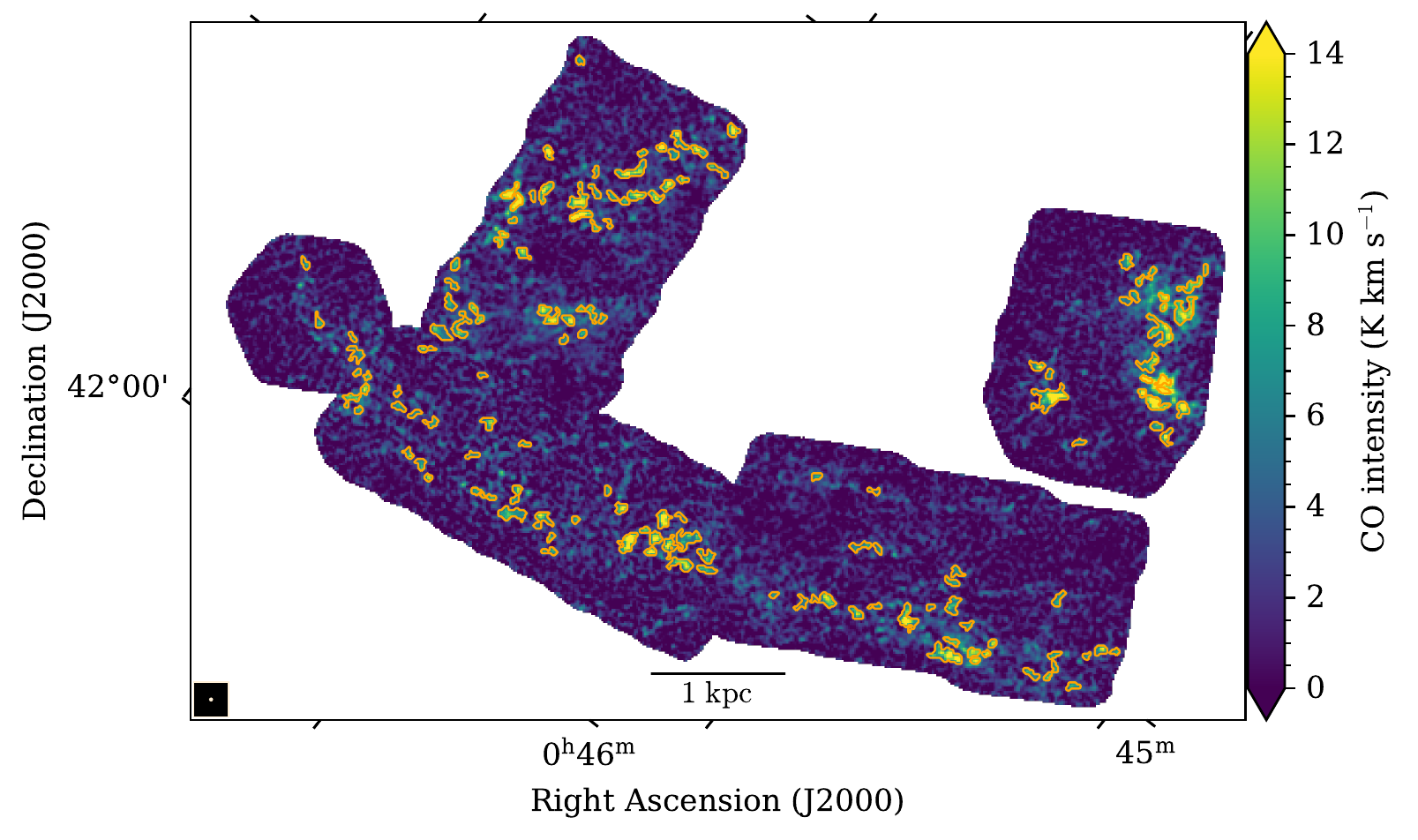}
\caption{Map of $^{12}$CO(J=1-0) intensity from CARMA + IRAM, convolved to 8" resolution and reprojected to 4" pixel size. Orange contours show the 140 sources extracted using a dendrogram.}
\label{fig:carma_cons_sources}
\smallskip
\end{figure*}

\begin{figure*}
\includegraphics[width=16.5cm]{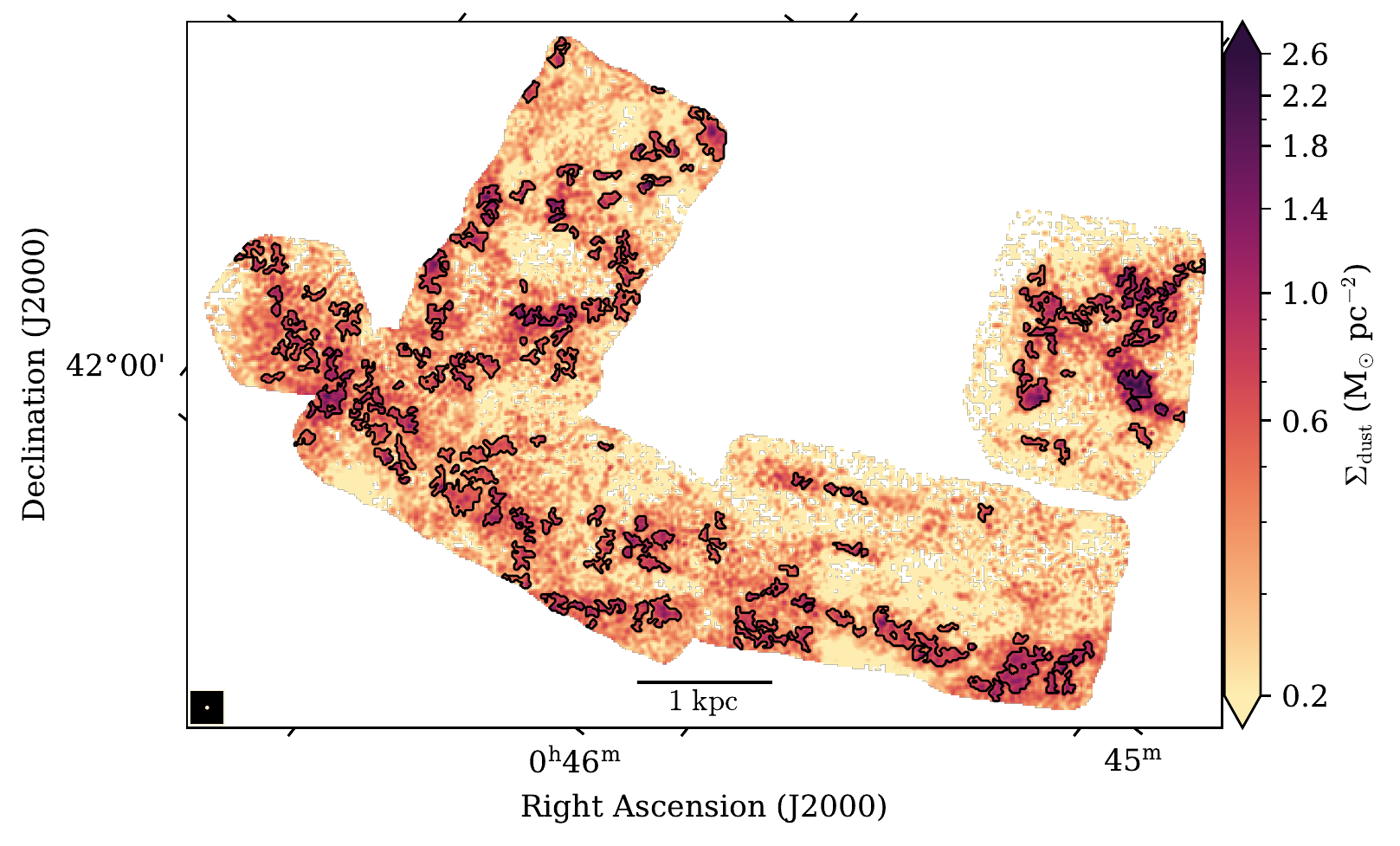}
\caption{Map of dust mass surface density produced by PPMAP at 8" resolution and 4" pixel size. Black contours show the 196 sources extracted using a dendrogram.}
\label{fig:dust_cons_sources}
\smallskip
\end{figure*}

 We also apply the dendrogram algorithm to our dust mass surface density map and identify sources (dendrogram `leaves') which we assume are molecular clouds traced by dust. Our motivation for finding dust-selected clouds stems from the existence of molecular gas in the MW which is not traced using the traditional CO method (see Section \ref{sec:intro}).

  Our dendrogram parameter selection is based on the work of the \textit{Planck} team (\citealt{PlanckCollaboration2011}) and allows us to extract a similar number of clouds as in our CO-selected catalogue. Figure \ref{fig:planck_cartoon} is an adaptation of Fig. 6 from \cite{PlanckCollaboration2011} showing the correlation between the optical depth of dust at a frequency of 857 GHz ($\tau_{857 \; \mathrm{GHz}}$) vs the combined atomic and molecular gas column density (N$_{\mathrm{H}}^{\mathrm{tot}}$) in the MW. The colour scale shows the number density of pixels on a logarithmic scale. At  N$_{\mathrm{H}}^{\mathrm{tot}}$ below $0.8 \times 10^{21}$ H cm$^{-2}$, the observed correlation between $\tau_{857 \; \mathrm{GHz}}$ and N$_{\mathrm{H}}^{\mathrm{tot}}$ (blue circles) follows a linear relationship in linear space (modelled by the red line). Above N$_{\mathrm{H}}^{\mathrm{tot}}$ $\approx$ 5 $\times$ 10$^{21}$ H cm$^{-2}$, N$_{\mathrm{H}}^{\mathrm{tot}}$ is dominated by contributions from CO, and $\tau_{857 \; \mathrm{GHz}}$ is visibly consistent with the observed linear correlation. An excess of $\tau_{857 \; \mathrm{GHz}}$ is seen between these two column density values (marked by the vertical red dashed lines) as the blue circles deviate from the linear relationship. This deviation from the modelled linear relationship is attributed by the \textit{Planck} team to CO-dark molecular gas.

  We select the optical depth of dust $\tau_{857 \; \mathrm{GHz}} = 3.0 \times 10^{-4}$ (see highest black dotted line in Figure \ref{fig:planck_cartoon}) to derive our input dendrogram parameters because it roughly corresponds to the upper limit of the deviation seen between the blue circles and the red line. This allows us to select a  \texttt{min\_value} below this (see lowest black dotted line in Figure \ref{fig:planck_cartoon}) and possibly pick up regions of CO-dark molecular gas. A frequency of 857 GHz roughly corresponds to a wavelength of 350 $\mu$m. We substitute our chosen $\tau_{857 \; \mathrm{GHz}} = 3.0 \; \times  10^{-4}$ and the dust mass absorption coefficient, $\kappa_{350, \; \mathrm{Draine}}$ = 0.192 m$^2$ kg$^{-1}$ (from \citealt{Draine2003}), into Equation \ref{eq:6} to find the corresponding $\Sigma_{\mathrm{dust}}$:
 \begin{equation}
 \label{eq:6}
 \begin{split}
 \Sigma_{\mathrm{dust}} = \frac{\tau_{857 \; \mathrm{GHz}}}{\kappa_{\mathrm{350, \; Draine}}} = \frac{3.0 \times 10^{-4}}{0.192 \; \mathrm{m^2} \; \mathrm{kg^{-1}}}  = 0.74 \; \mathrm{M_{\odot}} \; \mathrm{pc^{-2}}
 \end{split}
 \end{equation}

 For consistency with the 3:2 ratio of \texttt{min\_value} : \texttt{min\_delta} used to create our CO-selected catalogue, we split this $\Sigma_{\mathrm{dust}} = 0.74 \; \mathrm{M_{\odot}} \; \mathrm{pc}^{-2}$ value using the same ratio and obtain the \texttt{min\_value} and \texttt{min\_delta} for our dust-selected source extraction. The input dendrogram parameter values are listed in Table \ref{tab:param}.  Using these parameters, we find 196 clouds. Figure \ref{fig:dust_cons_sources} shows our dust-selected clouds. The spatial resolution of the \textit{Planck} observations ($\approx$ 0.3 pc; \citealt {PlanckCollaboration2011}) is, of course, much finer than our resolution. Therefore, our observations will not be as sensitive to CO-dark molecular gas as the \textit{Planck} observations of the MW.

 We calculate the total CO-traced molecular gas mass within our dust-selected clouds by finding the corresponding pixel locations in the CO map, and following the methodology described in Section \ref{ssec:gmc_co}. We also calculate the total dust mass of each cloud, CO-traced molecular GDR and total GDR as described in Section \ref{ssec:gmc_co}.

\section{Results}
\label{sec:res}
\subsection{Properties of extracted clouds}
\label{sec:cloud_mass}
We obtain two molecular cloud catalogues: one of clouds traced by CO and one of clouds traced by dust (Appendix \ref{sec:cloud_cats}). In this section, we provide our analysis of the cloud properties.

\begin{figure}
\centering
\includegraphics[width=8.5cm]{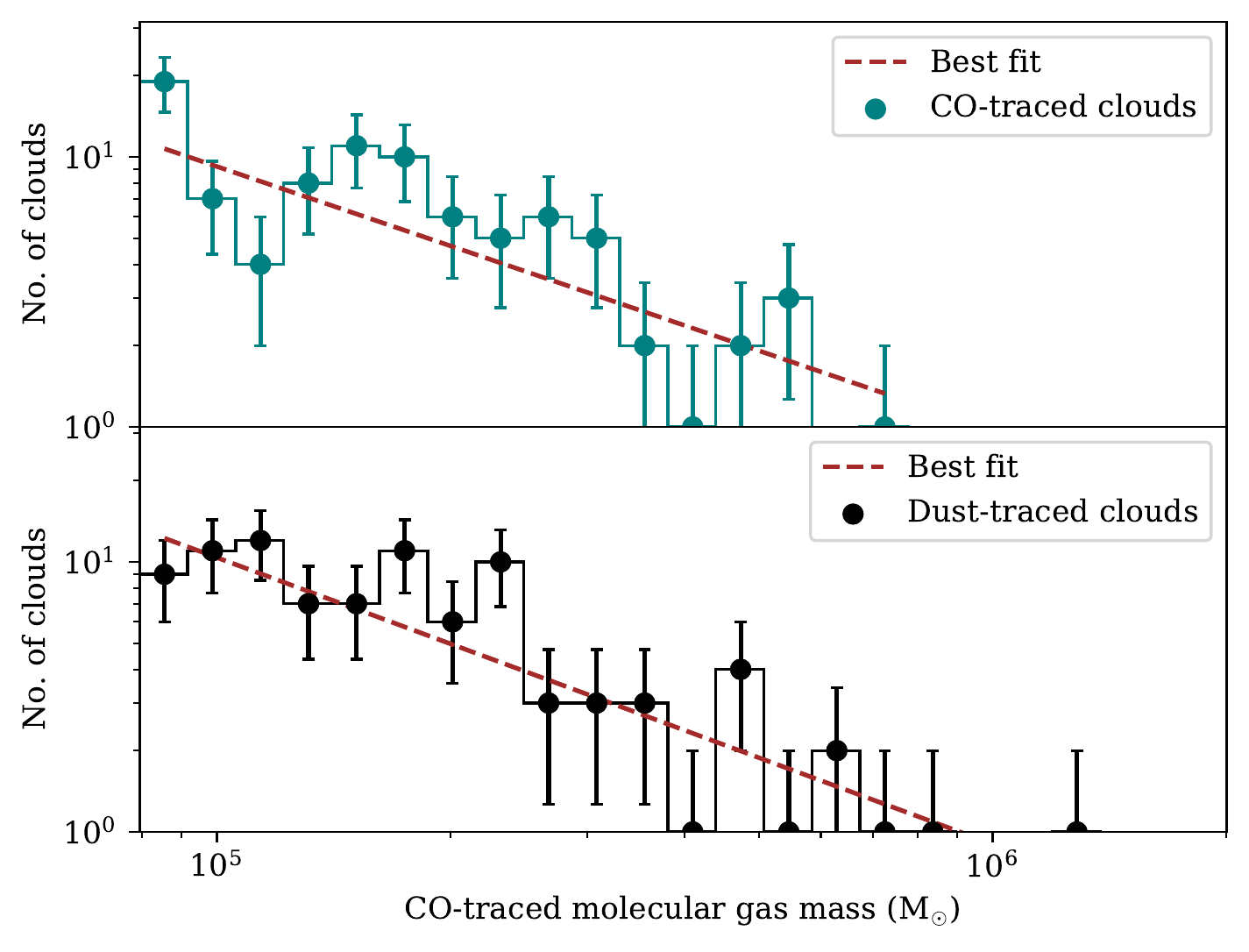} \caption{Cloud mass functions for CO-selected (teal) and dust-selected (black) catalogues, for cloud masses greater than 10$^{4.9}$ M$_{\odot}$. The histograms show the number of clouds per mass interval. The scatter points show the central mass value in each bin. The brown dashed line shows the best fit power law to the cloud mass function.}
\label{fig:cmf_all}
\end{figure}

Figure \ref{fig:cmf_all} shows the distribution of molecular cloud masses for our CO-selected and dust-selected catalogues. We fit a power law model to our cloud masses using the \texttt{SciPy} Levenberg-Marquardt least-squares minimisation package \texttt{lmfit}. We fit a power law of the form:
\begin{equation}
\label{eq:7}
  \mathrm{dN}_{\mathrm{cloud}} = \mathrm{k}M_{\mathrm{H_2}}^{-\alpha} \; dM_{\mathrm{H_2}}
\end{equation}

where dN$_{\mathrm{cloud}}$ is the number of molecular clouds in the mass interval ($M_{\mathrm{H_2}}$, $M_{\mathrm{H_2}}$ + $dM_{\mathrm{H_2}}$),  k is a normalisation factor, $M_{\mathrm{H_2}}$ is the total CO-traced molecular gas mass of the cloud, and $\alpha$ is the power law exponent.

Since the CO map is used to calculate the molecular gas mass of all dust-selected clouds, only the clouds with at least a 3$\sigma$ detection in the CO map have been included in our fits. 177 of 196 dust-traced clouds have a CO detection above this threshold. The molecular gas masses of the CO-selected clouds range from 3.9 $\times$ 10$^4$ M$_{\odot}$ $\le$ M$_{\mathrm{cloud}}$ $\le$ 7.1 $\times$ 10$^5$ M$_{\odot}$. The molecular gas masses of the dust-selected clouds with a 3$\sigma$ CO detection range from 1.8 $\times$ 10$^4$ M$_{\odot}$ $\le$ M$_{\mathrm{cloud}}$ $\le$ 1.3 $\times$ 10$^6$ M$_{\odot}$.
We only include clouds with a molecular gas mass $>10^{4.9}\ M_{\odot}= 7.9 \times 10^{4}\ M_{\odot}$ in our fitting because the decrease in the number of clouds at lower masses suggests that our mass functions are increasingly incomplete at lower masses. We bin data from both catalogues into 22 bins which are equidistant in logarithimic space between the mass limits 10$^{4.9}$ M$_{\odot}$ and 10$^{6.2}$ M$_{\odot}$. Our fits are performed in linear space and we assume Poisson errors.

\begin{table}
\caption{Cloud mass function best fit results and reduced $\chi^2$.}              
\label{tab:cmf_params}      
\centering                                      
\begin{tabular}{c c c c}         
\hline\hline                       
Data & Best fit $\alpha$ exponent & Reduced $\chi^2$ \\    
\hline                                  
     CO & 1.98 $\pm$ 0.24 & 1.38   \\      
     Dust & 2.06 $\pm$ 0.14 & 0.74   \\
\hline                                             
\end{tabular}
\end{table}

The best fit $\alpha$ values for clouds from each catalogue and the reduced $\chi^2$ parameter for our fits are given in Table \ref{tab:cmf_params}.
The reduced $\chi^2$ parameter for both catalogues is close to one, indicative of a good fit. The slope values of $\approx$ 2 are similar to measurements of the slope of the cloud mass function in the MW and nearby galaxies (e.g. \citealt{Rice2016}, \citealt{Rosolowsky2021}). They are also similar to the value for M31 by \cite{Kirk2015}: $\alpha$ = 2.34 $\pm$ 0.12.

Table \ref{tab:cprops} shows the 16th, 50th and 84th percentiles of cloud properties for both the CO-selected and the dust-selected clouds.

\begin{table*}
\caption{Statistical properties of clouds extracted from the $^{12}$CO(J=1-0) map and dust mass surface density map. Values have been rounded to 1 decimal place.}
\label{tab:cprops}      
\centering                                      
\begin{tabular}{l l l}          
\hline\hline                  
Properties & CO-selected & Dust-selected \\
\hline
No. of sources extracted & 140 & 196 \\
& & \\
50$^{84\mathrm{th}}_{16\mathrm{th}}$ percentile total CO-traced molecular gas mass in cloud & 9.4$^{23.0}_{6.0}$ $\times$ 10$^{4}$ M$_{\odot}$ & 7.1$^{20.6}_{2.6}$ $\times$ 10$^{4}$ M$_{\odot}$ \\
& & \\
50$^{84\mathrm{th}}_{16\mathrm{th}}$ percentile total dust mass in cloud & 2.8$^{7.0}_{1.8}$ $\times$ 10$^{3}$ M$_{\odot}$ & 5.5$^{10.4}_{2.4}$ $\times$ 10$^{3}$ M$_{\odot}$ \\
& & \\
50$^{84\mathrm{th}}_{16\mathrm{th}}$ percentile density weighted average $T_{\mathrm{dust}}$ in cloud & 15.6$^{17.1}_{14.3}$ K & 14.6$^{15.9}_{13.8}$ K \\
& & \\
50$^{84\mathrm{th}}_{16\mathrm{th}}$ percentile density weighted average $\beta$ in cloud & 2.1$^{2.3}_{2.0}$ & 2.2$^{2.5}_{2.0}$ \\
& & \\
50$^{84\mathrm{th}}_{16\mathrm{th}}$ percentile molecular GDR of cloud & 34.5$^{48.3}_{25.8}$ & 14.1$^{25.8}_{7.2}$ \\
& & \\
50$^{84\mathrm{th}}_{16\mathrm{th}}$ percentile total GDR of cloud & 80.0$^{107.6}_{54.5}$ & 55.1$^{70.1}_{34.9}$ \\
& & \\
50$^{84\mathrm{th}}_{16\mathrm{th}}$ percentile equivalent radius of cloud & 36.9$^{50.8}_{29.7}$ pc & 47.8$^{68.7}_{33.3}$ pc \\
& & \\
\hline                                   
\end{tabular}
\end{table*}

\subsection{Radial variations in dust properties of clouds traced by CO}
\label{ssec:radvarres}
\begin{figure*}
\centering
\includegraphics[width=13cm]{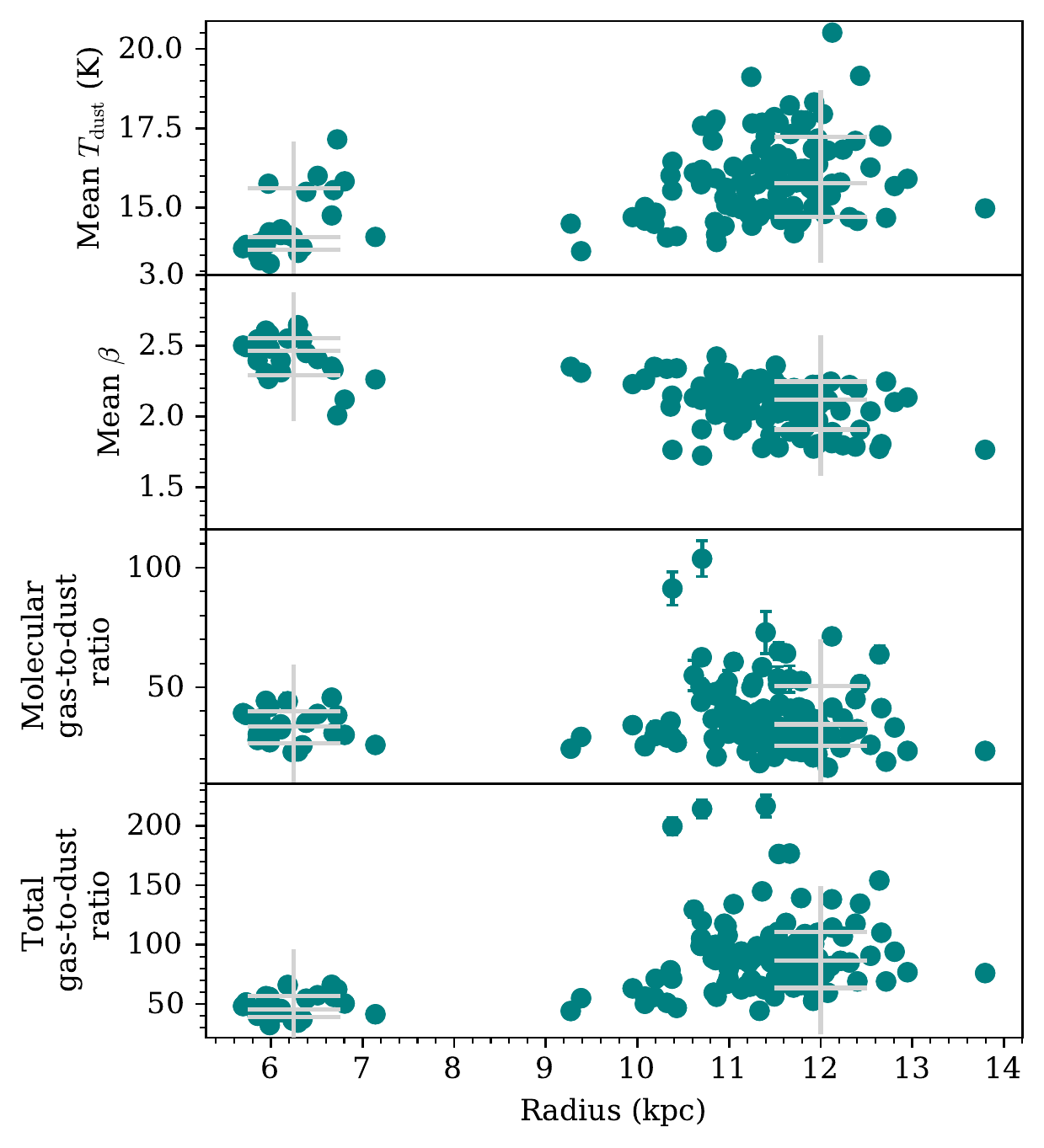} \caption{Radial distribution of gas and dust properties of clouds from the CO-selected catalogue. Each scatter point represents a dendrogram leaf (molecular cloud). The light grey crosses show the 16th, 50th and 84th percentiles of each cloud property in the 5-7.5 kpc radial bin and the 9-15 kpc radial bin. \textit{First row:} Radial distribution of mean $T_{\mathrm{dust}}$ in each cloud. \textit{Second row:} Radial distribution of mean $\beta$ in each cloud. \textit{Third row:} Radial distribution of CO-traced molecular GDR of each cloud. \textit{Fourth row:} Radial distribution of total (H{\sc i} + H$_2$) GDR of each cloud. The GDR errorbars have been calculated using the method described in Section \ref{sssec:gdpropmethod}.}
\label{fig:properties}
\end{figure*}

Figure \ref{fig:properties} shows the mean $T_{\mathrm{dust}}$, mean $\beta$, CO-traced molecular GDR, and the total GDR as a function of radius for clouds from the CO-selected catalogue. We find that our extracted clouds are located within two
ranges of radius: between 5-7.5 kpc and between 9-15 kpc, reflecting the
inner ring at $\approx$ 5 kpc and the dusty star-forming ring at $\approx$ 10 kpc.

\begin{figure*}
\centering
\includegraphics[width=18cm]{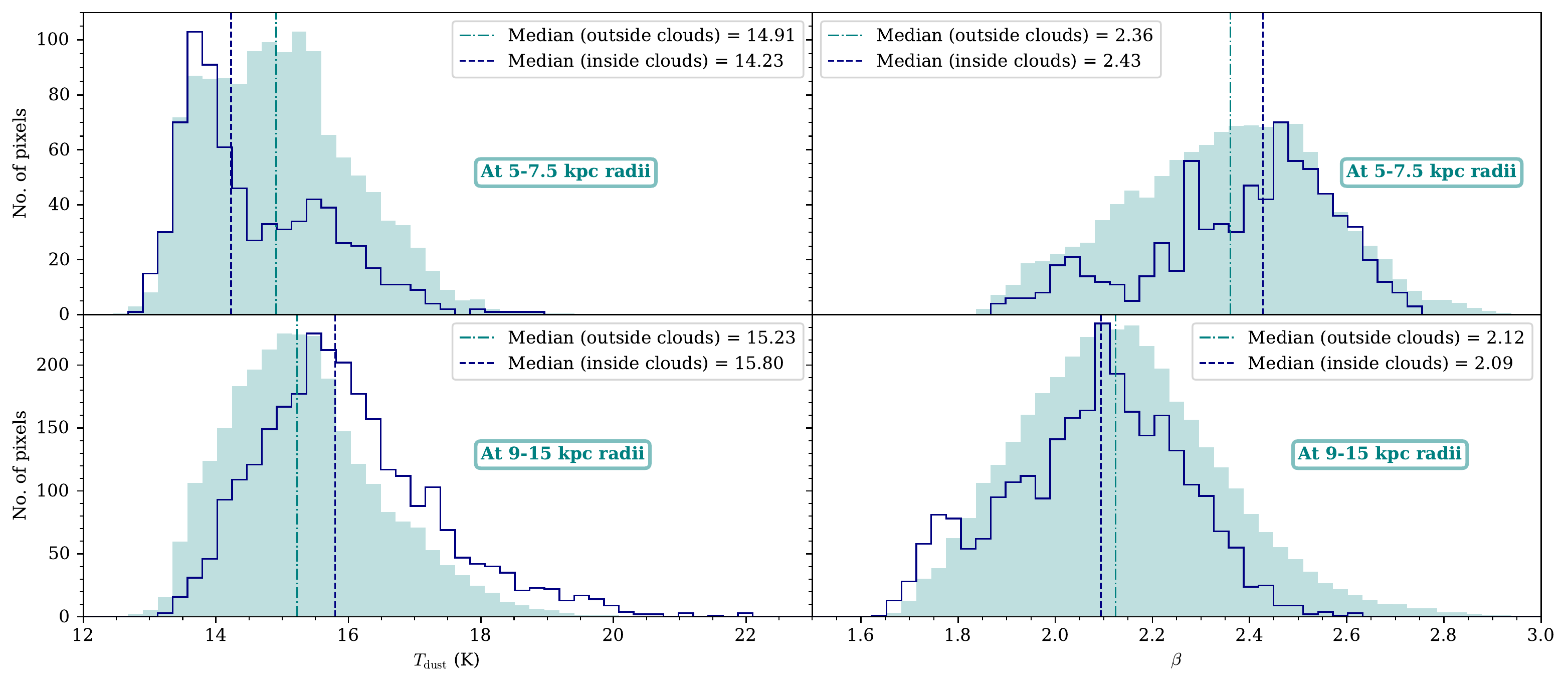} \caption{Distribution of $T_{\mathrm{dust}}$ and $\beta$ inside (navy) and outside molecular clouds (shaded teal) from the CO-selected catalogue. \textit{Top left:} $T_{\mathrm{dust}}$ distribution for pixels in the radius range 5-7.5 kpc. \textit{Top right:} $\beta$ distribution for pixels in the radius range 5-7.5 kpc. \textit{Bottom left:} $T_{\mathrm{dust}}$ distribution for pixels in the radius range 9-15 kpc. \textit{Bottom right:} $\beta$ distribution for pixels in the radius range 9-15 kpc. In all four plots, the histogram of pixels outside clouds has been rescaled to have the same height as the histogram of pixels inside clouds. The navy dashed vertical line shows the median value of $T_{\mathrm{dust}}$ and $\beta$ inside clouds. The teal dashed-dotted vertical line shows the median value outside clouds.}
\label{fig:tempbetaCARMAvar}
\end{figure*}

Figure \ref{fig:tempbetaCARMAvar} shows the mean $T_{\mathrm{dust}}$ and
mean $\beta$ in these two radial bins, both for the pixels inside and outside clouds.
The left-hand column of Figure \ref{fig:tempbetaCARMAvar} shows histograms of the $T_{\mathrm{dust}}$. We see that the median $T_{\mathrm{dust}}$ increases with galactocentric radius. The distributions of $T_{\mathrm{dust}}$ for pixels inside clouds but at different radii are significantly different (two-sided Mann-Whitney U test, p-value $\ll$ 0.01; see Table \ref{tab:mannwhitneyu} for full results), as are the distributions of $T_{\mathrm{dust}}$ for the pixels outside clouds (p-value $\ll$ 0.01). However, the temperature difference is fairly small. The average $T_{\mathrm{dust}}$ is 1.57 K higher for the clouds in the 10 kpc ring than for the clouds in the 5 kpc ring, and the average $T_{\mathrm{dust}}$ is 0.32 K higher for the pixels outside the clouds in the 10 kpc ring than for the pixels outside the clouds in the 5 kpc ring.
In both radial bins, the distributions of $T_{\mathrm{dust}}$ inside and outside molecular clouds are significantly different (p-value $\ll$ 0.01), although again the difference in temperature is actually quite small, $\simeq 0.6-0.7$ K.

The right-hand column of Figure \ref{fig:tempbetaCARMAvar} shows our most interesting result, with histograms of $\beta$ in pixels inside and outside molecular clouds, split into the two radial bins. The distributions of $\beta$ for pixels inside clouds at different radii are significantly different (two-sided Mann-Whitney U test, p-value $\ll$ 0.01; see Table \ref{tab:mannwhitneyu} for full results), as are the distributions of $\beta$ for the pixels outside clouds (p-value $\ll$ 0.01). The average $\beta$ is 0.34 lower for the clouds in the 10 kpc ring than for the clouds in the 5 kpc ring, and the average $\beta$ is 0.24 lower for the pixels outside the clouds in the 10 kpc ring than in the 5 kpc ring. Our result is consistent with the decreasing trend in $\beta$ with increased galactocentric radius (going from $\beta \approx$ 2.5 to $\beta \approx$ 1.9) discovered by \cite{Smith2012} beyond a radius of 3.1 kpc. Our result is also in agreement with the results from
other \textit{Herschel} observations of M31 by \cite{Draine2014}, which found larger $\beta$ in the central $\approx$ 7 kpc than in the outer disk (see their Fig. 13). Although \cite{Draine2014} find a shallower decrease in $\beta$ beyond $\gtrsim$ 7 kpc than \cite{Smith2012}, our result agrees with the general trend found by both studies of a radial decrease in $\beta$ between the radii of 5-15 kpc.

 In both radial bins, the distributions of $\beta$ inside and outside molecular clouds are significantly different (p-value $\ll$ 0.01). However, the more interesting aspect of our result is that there is a much smaller difference between the average $\beta$ inside molecular clouds compared to outside molecular clouds in both radial bins. The median value of $\beta$ is only greater by 0.07 inside clouds in the 5 kpc ring and is actually less by 0.03 inside the clouds in the 10 kpc ring.
 Therefore, we find no evidence for the radial variations in $\beta$ in M31 being caused by a large change in $\beta$ in regions of dense gas.

 \sisetup{
     tight-spacing           = true,
     round-mode              = places,
     round-precision         = 1,
     scientific-notation     = true,
     }
 \begin{table*}
 \caption{Two-sided Mann-Whitney U test results for $T_{\mathrm{dust}}$ and $\beta$ inside and outside molecular clouds at the inner and outer ring; and distributions of molecular GDR and total GDR for CO-traced and dust-traced clouds with at least a 3$\sigma$ detection.}              
 \label{tab:mannwhitneyu}      
 \centering                                      
 \begin{tabular}{c c S S}          
 \hline\hline                        
 Samples being compared & Sample sizes (no. of pixels/clouds) & {U-statistic} & {p-value} \\    
\hline
      $T_{\mathrm{dust}}$ inside clouds in inner ring vs outer ring & Inner ring = 734; Outer ring = 2672  & 396546.0 & \num{1.2662890024612077e-31} \\                               
      $T_{\mathrm{dust}}$ outside clouds in inner ring vs outer ring & Inner ring = 9807; Outer ring = 50826 & 205272680.0 & \num{7.953400849866134e-169} \\      
      $T_{\mathrm{dust}}$ inside vs outside clouds in the inner ring & Inside clouds =734 ; Outside clouds = 9807 & 4415974.5 & \num{9.477798817217171e-25} \\
      $T_{\mathrm{dust}}$ inside vs outside clouds in the outer ring & Inside clouds = 2672; Outside clouds = 50826 & 48974193.5 & \num{1.0107338150253324e-130} \\
      \hline
      $\beta$ inside clouds in inner ring vs outer ring & Inner ring = 734; Outer ring = 2672 & 1713391.0 & \num{1.038490038747912e-211} \\
      $\beta$ outside clouds in inner ring vs outer ring & Inner ring = 9807; Outer ring = 50826 & 384633211.5 & {$\ll$ 0.01} \\
      $\beta$ inside vs outside clouds in the inner ring & Inside clouds = 734 ; Outside clouds = 9807 & 3111901.5 & \num{8.929688018379722e-10} \\
      $\beta$ inside vs outside clouds in the outer ring & Inside clouds = 2672; Outside clouds = 50826 & 77007902.5 & \num{1.2662890024612077e-31} \\
 \hline                                             
      Molecular GDR for CO-traced vs dust-traced clouds & CO-traced clouds = 140; Dust-traced clouds = 177 & 22854.0 & \num{3.836995338497158e-38} \\
      Total GDR for CO-traced vs dust-traced clouds & CO-traced clouds = 140; Dust-traced clouds = 177 & 19146.0 & \num{7.65198820850759e-17} \\
 \hline
 \end{tabular}
 \end{table*}

\subsection{The gas-to-dust ratio in clouds found using the two methods}
\label{ssec:gdrres}

The left-hand column of Figure \ref{fig:gdr_mass_cat} shows histograms of the CO-traced molecular GDR of clouds from both of our cloud catalogues. The median molecular GDR of clouds from the dust-selected catalogue is roughly half the value for the CO-selected
catalogue. We compare the two distributions using a Mann-Whitney U test,
finding that the distributions are significantly different (p-value $\ll$ 0.01; see Table \ref{tab:mannwhitneyu} for full results).
The right-hand column of Figure \ref{fig:gdr_mass_cat} shows histograms of the total (H{\sc i} + H$_2$) GDR of clouds from both of our cloud catalogues. We compare the two distributions using the Mann-Whitney U test, again finding a significant difference (p-value $\ll$ 0.01), with the
GDR for the dust-selected clouds being lower than for the CO-selected
clouds.

\section{Discussion}
\label{sec:disc}

\subsection{Dust emissivity index inside and outside molecular clouds}
\label{ssec:betavarydisc}
\cite{Smith2012} have found radial variations in $\beta$ in M31, with the value of $\beta$ decreasing from $\simeq$ 2.5 at a radius of 3 kpc to $\simeq$ 1.8 at 12 kpc. \cite{Draine2014} have also found, using a different \textit{Herschel} dataset and a different method, the same general trend; with $\beta$ decreasing from a value of $\simeq 2.35$ at a radius of 3 kpc to a value of $\simeq$ 2.0 at a radius of 12 kpc. Some possible causes of these variations are large grain coagulation or the accretion of a mantle in denser environments since some dust models (e.g. \citealt{Kohler2015}) predict that this will lead to an increase in the value of $\beta$ by $0.3-0.5$.

In our study, we find a much smaller difference (of order 0.03 to 0.07) between the median values of $\beta$ in low-density and high-density environments at the same
radius than the much larger radial change. Our results are similar to the findings of \cite{Roman-Duval2017a} in the Magellenic Clouds, who find no correlation between $\beta$ and gas mass surface density at spatial scales of 75 pc in the Large Magellenic Cloud (LMC) and 90 pc in the Small Magellenic Cloud (SMC). Our result contrasts with the increase in $\beta$, found along sight lines dominated by molecular gas, in the MW by \cite{PlanckCollaboration2014a}. The \textit{Planck} team find that $\beta$ increases from 1.75 in the atomic medium of the Galactic plane to 1.98 in the molecular medium.  Our result also contrasts with what has been found in M33 where $\beta$ (from SED fitting of \textit{Herschel} observations at $\approx$ 150 pc spatial scales) is strongly and positively correlated with molecular gas traced by $^{12}$CO(J=2-1) emission (\citealt{Tabatabaei2014}).
The authors do not attempt to separate effects of radius on $\beta$ from the effects of a dense environment in their study. Therefore, the apparent increase in $\beta$ in dense environments in M33 might be caused by a radial gradient that is unconnected to gas density
but caused by there being more clouds at small radii in the galaxy.
We note that our study is the first one that has tried to
separate the effect of gas density and radius on $\beta$ in M31, although we examine a smaller dynamic range in galactocentric radii.

To explore the possibility that our results
are the consequence of our choice of dendrogram parameters, we have run the dendrogram with a set of parameters that resulted in a peak CO threshold of 3$\sigma$ rather than 5$\sigma$. We found that our results
were very similar.

Our strong conclusion therefore is that in M31, molecular gas surface density is not the driver of radial variations in $\beta$ at 30 pc spatial scales. This suggests that, at these spatial scales, large grain coagulation in dense environments is not having a big effect on $\beta$. The only alternative is that there is some genuine radial change in the composition or the structure of the dust grains as we move out through the galaxy. What these changes are is still a mystery, although one speculative possibility is that there is a radial change in the ratio of carbonaceous and silicate dust grains, perhaps caused by a changing C/Si abundance ratio.

\subsection{CO-dark gas?}
\label{ssec:dustyismdisc}

\begin{figure*}
\centering
\hspace{-0.4cm}
\includegraphics[width=16.5cm]{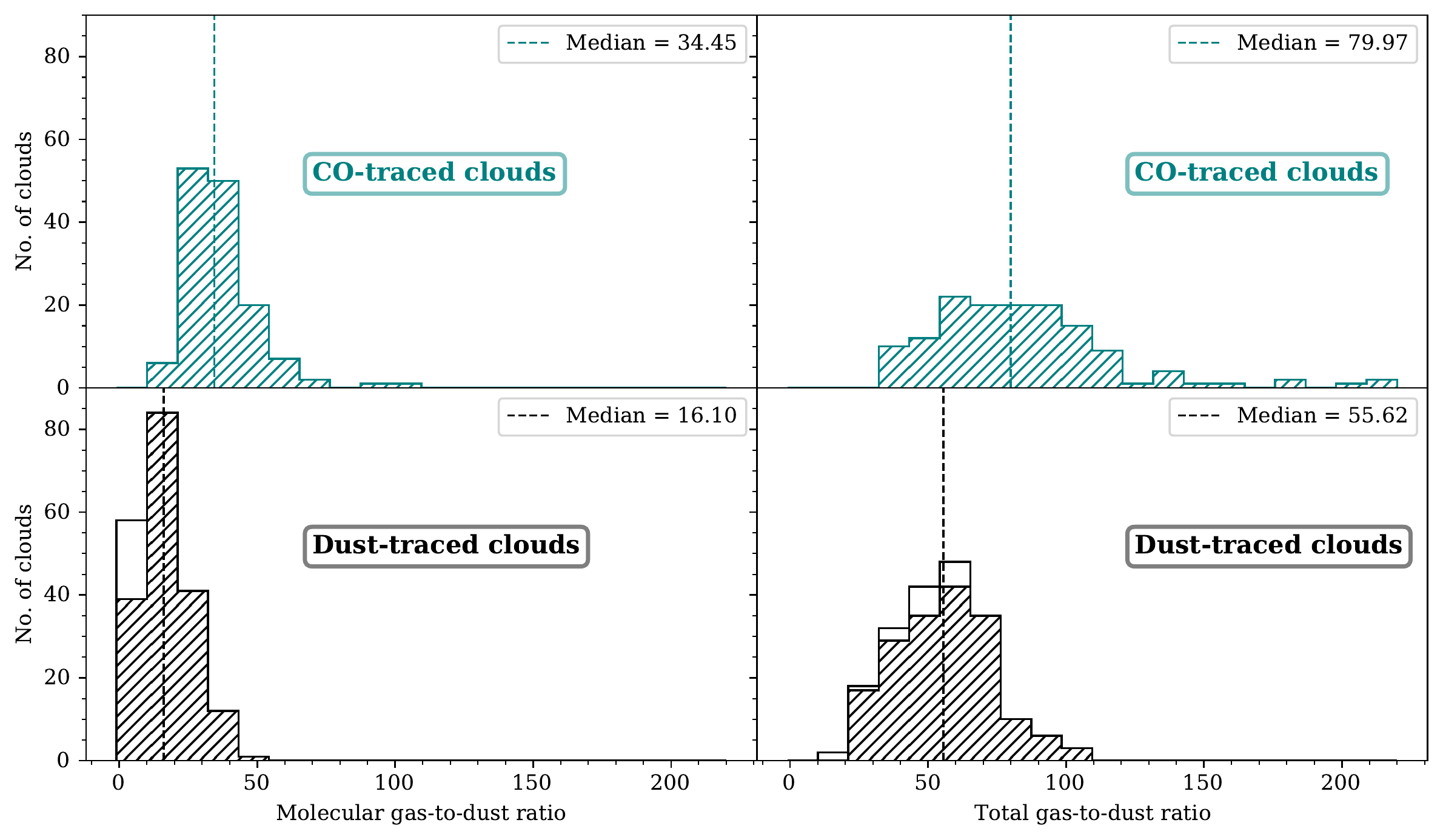} \caption{Histograms showing the number of clouds from the CO-selected and dust-selected catalogues and their CO-traced molecular GDR and total GDR. \textit{Top:} Clouds from the CO-selected catalogue. \textit{Bottom:} Clouds from the dust-selected catalogue. The hatched regions indicate clouds with a GDR measurement and the non-hatched regions represent an upper limit. The vertical dashed lines indicate the median GDR values for each hatched histogram.}
\label{fig:gdr_mass_cat}
\end{figure*}

\begin{figure*}
\includegraphics[width=16.5cm]{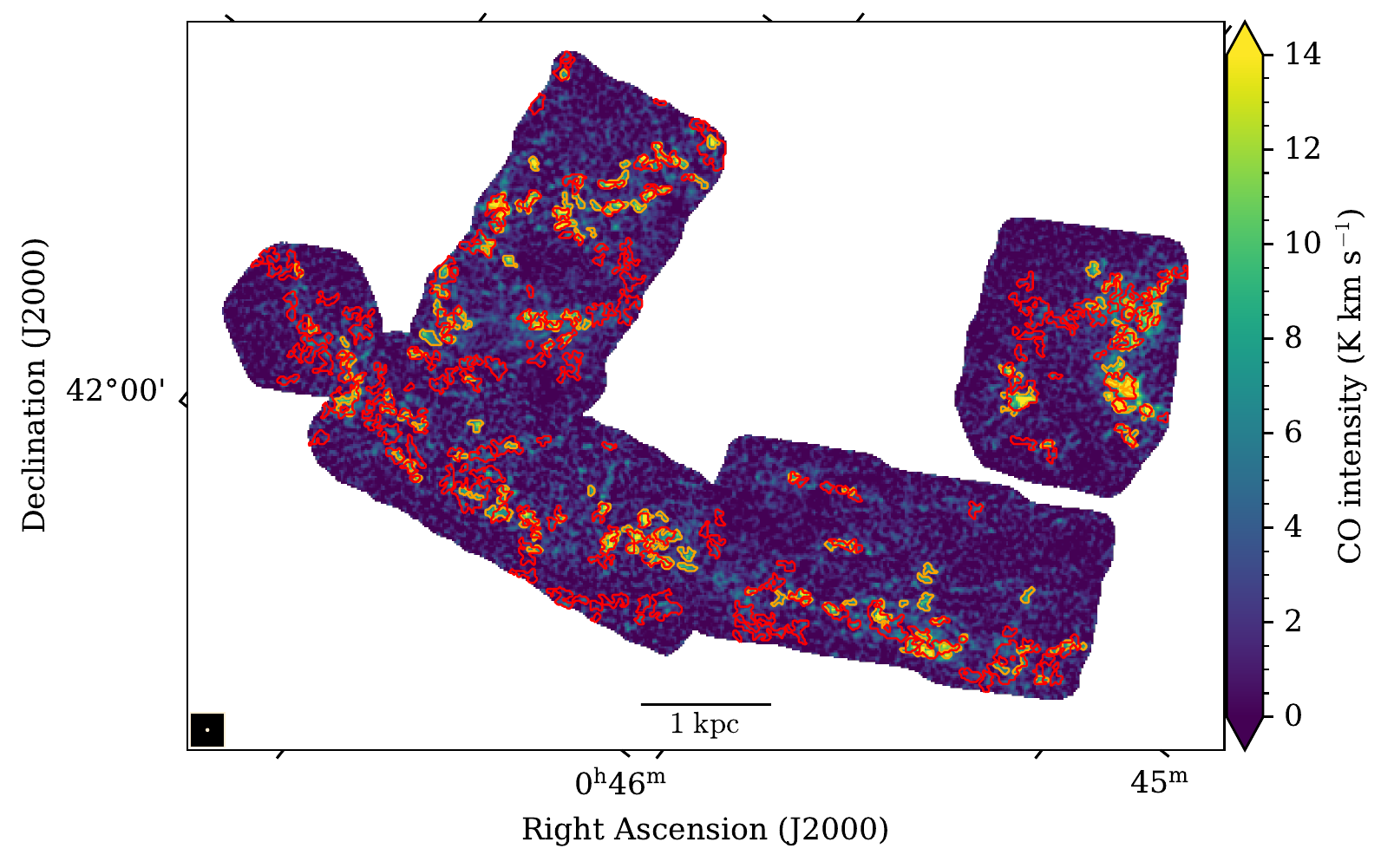}
\caption{Map of $^{12}$CO(J=1-0) intensity taken from CARMA + IRAM, convolved to 8" resolution and reprojected to 4" pixel size. Orange contours show the 140 sources from the CO-selected catalogue. The red contours show the sources from the dust-selected catalogue.}
\label{fig:comap_overlap}
\smallskip
\end{figure*}

\begin{figure*}
\includegraphics[width=16.5cm]{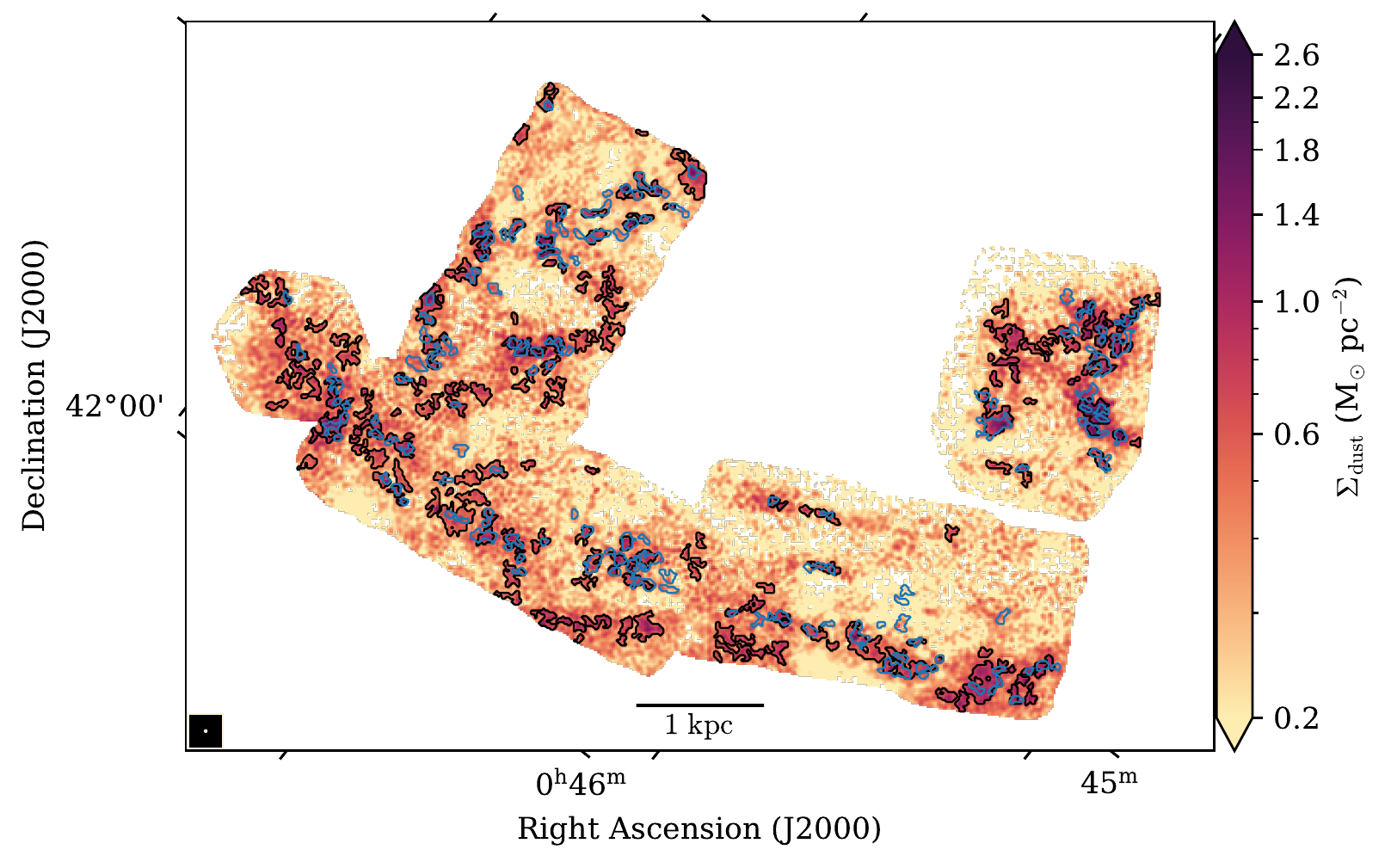}
\caption{Map of dust mass surface density produced by PPMAP at 8" resolution and 4" pixel size. Black contours show the 196 sources from the dust-selected catalogue. The blue contours show the sources from the CO-selected catalogue.}
\label{fig:dustmap_overlap}
\smallskip
\end{figure*}

In Section \ref{ssec:gdrres}, we have compared the GDR distributions of clouds from the two catalogues,
finding that the GDR for the dust-selected clouds is significantly lower than for the CO-selected clouds. Figure \ref{fig:gdr_mass_cat} shows that there are clouds in the dust-selected catalogue with molecular GDR below $\approx$ 16, which are not found in the CO-selected catalogue. Figures \ref{fig:comap_overlap} and \ref{fig:dustmap_overlap} show the overlap of our CO-selected and dust-selected clouds. Alongside dust-selected clouds with low CO emission, we serendipitously find some CO-selected clouds with low dust emission. We are uncertain of the astrophysical meaning of these clouds.

One possible explanation for clouds in the dust-selected catalogue with low molecular GDR is that they may simply be emphemeral structures in atomic gas. We have investigated this by adding in the atomic gas mass and finding the total (H{\sc i} + H$_2$) GDR of clouds (see right-hand column of Figure \ref{fig:gdr_mass_cat}). However, even after adding in the atomic gas mass, we find some dust-selected clouds with lower total GDR than any of the CO-selected clouds.

Are the clouds showing lower levels of CO simply because they are smaller, which might suggest that they are not genuine molecular cloud structures? To answer this question, we have compared the physical sizes and total masses of the clouds in our dust-selected
catalogue. We have found the total ISM mass of each cloud by multiplying the dust mass surface density of our cloud by a constant GDR of 100 (\citealt{Hildebrand1983}) and calculating the total dust mass of the cloud as described in Section \ref{ssec:gmc_ppmap}.
We have used the dust mass rather than the CO to estimate the total ISM mass because we want to examine the possibility that there are molecular clouds made up of a large proportion of CO-dark gas.
As for the physical size, we have simply taken the area of each cloud in square parsec.

\begin{figure*}
\centering
\includegraphics[width=13cm]{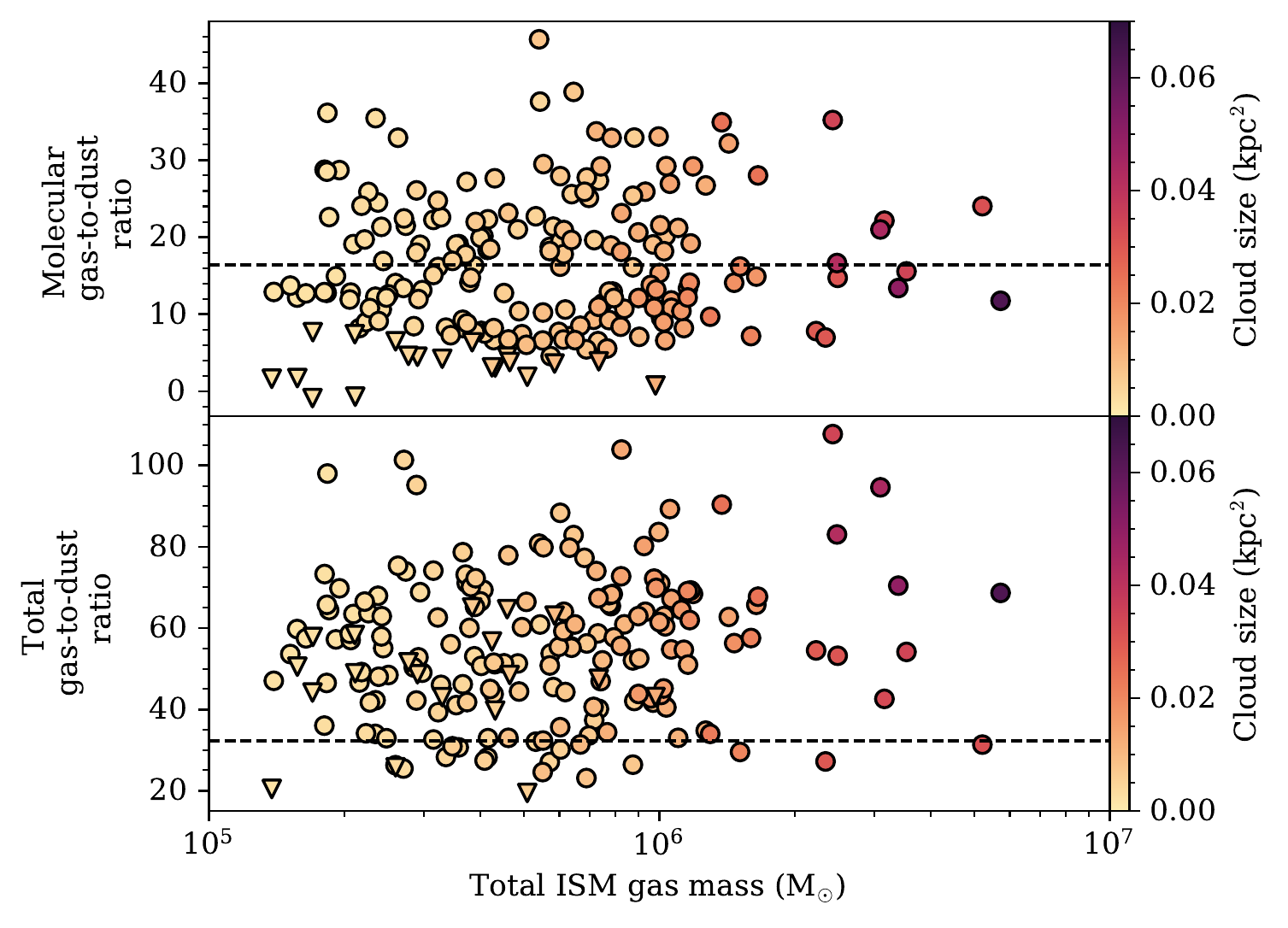} \caption{\textit{Top row:} Total ISM mass (inferred from the dust) versus CO-traced molecular GDR of dust-selected clouds. \textit{Bottom row:} Total ISM mass (inferred from dust) versus total GDR of dust-selected clouds. The points are coloured by cloud size given in terms of its physical area in kpc$^2$. The circles represent dust-selected clouds with at least a 3$\sigma$ detection in the CO map. The triangles represent dust-selected clouds with less than a 3$\sigma$ detection in the CO map. The dashed horizontal black line shows the lowest molecular GDR obtained (GDR $\approx$ 16) and lowest total GDR obtained (GDR $\approx$ 32) in clouds from the CO-selected catalogue.}
\label{fig:gdr_mass_cloudsize}
\end{figure*}

The top panel of Figure \ref{fig:gdr_mass_cloudsize} shows the total ISM mass versus the molecular GDR in the dust-selected clouds coloured by cloud size. The bottom panel shows the total ISM mass vs total GDR of clouds. We find that there is still a population of large clouds ($\geq$ 0.03 kpc$^2$) with ISM mass greater than 10$^6$ M$_{\odot}$ and low total GDR ($\lesssim$ 50). There are some clouds with total GDR below $\approx$ 32 which are not found in the CO-selected catalogue.

One possibility is that these are real molecular clouds with low levels of CO, i.e. clouds that are largely made up of CO-dark gas. The alternative explanation remains that these structures are largely made up of atomic gas, perhaps the result of source confusion along the line of sight. Although we have tried to correct for the contribution of atomic gas,
we are limited by the resolution of the available radio data.
The HARP and SCUBA-2 High-Resolution Terahertz Andromeda Galaxy (HASHTAG) survey currently underway at the James Clerk Maxwell Telescope, which will give a higher resolution view of the dust, and a new Very Large Array (VLA) survey, which will give a higher resolution view of the atomic gas, will help to distinguish between these two possibilities.

\section{Summary}
\label{sec:conc}

M31 forms an excellent testbed for understanding the variations in dust properties and the interplay of dust and gas in a spiral galaxy similar to our own. We investigate whether radial variations in the dust emissivity index ($\beta$) are caused by an increase of $\beta$ in dense molecular gas regions in M31. We probe the ISM of M31 at significantly improved spatial resolution ($\approx$ 30 pc) compared to previous studies, using combined CARMA + IRAM $^{12}$CO(J=1-0) observations and {\it Herschel} observations of the dust which have been reanalysed using the PPMAP algorithm.
We use a dendrogram to create molecular cloud catalogues in two ways: using CO and dust as a tracer. Our key findings are:

 \begin{enumerate}
    \item We see a radial variation in $\beta$ in agreement with previous studies (e.g. \citealt{Smith2012}, \citealt{Draine2014}, \citealt{Whitworth2019}), with a decrease in $\beta$ going from the inner ring to outer dusty, star-forming ring.
    \item We find no evidence for radial variations in $\beta$ being caused by an increase of $\beta$ in dense molecular gas regions at radii between $5-7.5$ kpc and $9-15$ kpc.
    \item We find a population of clouds in our dust-selected catalogue with lower median CO-traced molecular GDR than in our CO-selected catalogue. These may be
    clouds containing CO-dark molecular gas, although we are unable to rule
    out the possibility that these structures are confused with features
    in the atomic phase of the ISM.
 \end{enumerate}

 We conclude that an increase of $\beta$ in dense molecular gas regions is not the prominent driver of the radial variations in $\beta$ in M31.

\section*{acknowledgements}
    GAE acknowledges the support of a UK Science and Technology Facilities Council (STFC) postgraduate studentship. GAE personally thanks Michael Anderson for useful discussions about the dendrogram algorithm. SAE, MWLS and APW acknowledge the support of a STFC Consolidated Grant (ST/K00926/1). MWLS acknowledges support from the European Research Council (ERC) in the form of the Consolidator Grant {\sc CosmicDust} (ERC-2014-CoG-647939, PI H\,L\,Gomez). We thank the anonymous referee for suggestions which helped improved the quality of the manuscript. This research made use of \texttt{astropy}\footnote{\texttt{astropy} webpage: \url{https://www.astropy.org}}, \texttt{cmocean}\footnote{\texttt{cmocean} webpage: \url{https://matplotlib.org/cmocean/}},  \texttt{matplotlib}\footnote{\texttt{matplotlib} webpage: \url{https://matplotlib.org}} (\citealt{Hunter2007}), \texttt{multicolorfits}\footnote{\texttt{multicolorfits} GitHub page: \url{https://github.com/pjcigan/multicolorfits}}, \texttt{numpy}\footnote{\texttt{numpy} webpage: \url{https://numpy.org}} (\citealt{Harris2020}) and \texttt{scipy}\footnote{\texttt{scipy} webpage: \url{https://www.scipy.org/index.html}} (\citealt{SciPy1.0Contributors2020}) Python packages and the  \texttt{iPython}\footnote{\texttt{iPython} webpage: \url{https://ipython.org}} (\citealt{Perez2007}) interactive shell. This research has made use of SAOImageDS9\footnote{SAOImageDS9 webpage: \url{https://sites.google.com/cfa.harvard.edu/saoimageds9/download}} (\citealt{Joye2003}) astronomical imaging and data visualisation software.

\section*{Data Availability}
The raw far-infrared data used for this work are available on the Herschel Science Archive\footnote{Herschel Science Archive: \url{http://archives.esac.esa.int/hsa/whsa/?ACTION=PUBLICATION&ID=2019MNRAS.489.5436W}}. The raw $^{12}$CO(J=1-0) data used for this work are available on the CARMA data archive. The raw HI data used for this work are available upon request from Robert Braun.



\bibliographystyle{mnras}
\bibliography{Dust_Emissivity_Index_in_M31_revision_3} 




\begin{appendix}

\section{Cloud catalogues}
\label{sec:cloud_cats}

\begin{landscape}
\setlength\tabcolsep{9pt}

\sisetup{
    tight-spacing           = true,
    round-mode              = places,
    round-precision         = 1,
    }

\begin{table}
  \label{tab:COcat}
\caption{Sample of 10 clouds from the CO-selected catalogue, ordered by peak CO intensity, and their properties. The full catalogue, including the cloud sizes, is available as supplementary material.}
\begin{tabular}{cccSSccccSc}
\hline \hline
Cloud ID & Mean $T_{\mathrm{dust}}$ & Mean $\beta$ & {Total CO-traced gas mass} & {Total dust mass} & CO-traced GDR & RA [J2000] & Dec [J2000] & Radius to peak & {Total gas mass} & Total GDR \\
$\mathrm{}$ & ($\mathrm{K}$) & $\mathrm{}$ & {($\mathrm{M_{\odot}}$)} & {($\mathrm{M_{\odot}}$)} & $\mathrm{}$ & ($\mathrm{{}^{\circ}}$) & ($\mathrm{{}^{\circ}}$) & ($\mathrm{kpc}$) & {($\mathrm{M_{\odot}}$)} & $\mathrm{}$ \\
\hline
22 & 18.3 & 2.1 & 4.1E+05 & 1.1E+04 & 36.7 & 11.3 & 41.6 & 11.9 & 7.1E+05 & 100.6 \\
93 & 14.0 & 2.3 & 2.0E+05 & 6.0E+03 & 32.7 & 11.1 & 41.6 & 5.9 & 4.2E+04 & 39.6 \\
96 & 13.2 & 2.5 & 1.8E+05 & 6.5E+03 & 27.0 & 11.1 & 41.6 & 6.0 & 3.4E+04 & 32.2 \\
44 & 15.6 & 2.0 & 4.5E+05 & 7.0E+03 & 64.2 & 11.4 & 41.8 & 11.6 & 3.8E+05 & 118.2 \\
95 & 13.7 & 2.4 & 2.9E+05 & 8.1E+03 & 35.6 & 11.0 & 41.6 & 5.9 & 6.2E+04 & 43.2 \\
179 & 15.8 & 1.8 & 1.7E+05 & 2.4E+03 & 71.3 & 11.2 & 41.9 & 12.1 & 1.6E+05 & 138.1 \\
160 & 16.7 & 2.2 & 2.8E+05 & 5.6E+03 & 51.2 & 11.2 & 41.9 & 11.5 & 1.8E+05 & 83.4 \\
184 & 17.9 & 2.3 & 1.3E+05 & 3.7E+03 & 34.5 & 11.1 & 41.9 & 11.5 & 9.7E+04 & 60.5 \\
78 & 15.8 & 2.3 & 2.6E+05 & 8.8E+03 & 29.1 & 11.1 & 41.6 & 6.0 & 1.3E+05 & 43.7 \\
158 & 15.6 & 2.3 & 3.0E+05 & 8.0E+03 & 38.0 & 11.2 & 41.9 & 11.0 & 2.3E+05 & 66.8 \\
\end{tabular}
\end{table}

\begin{table}
  \label{tab:dustcat}
\caption{Sample of 10 clouds from the dust-selected catalogue, ordered by peak dust mass surface density, and their properties. The full catalogue, including the cloud sizes, is available as supplementary material.}
\begin{tabular}{cccSSccccSc}
\hline \hline
Cloud ID & Mean $T_{\mathrm{dust}}$ & Mean $\beta$ & {Total CO-traced gas mass} & {Total dust mass} & CO-traced GDR & RA [J2000] & Dec [J2000] & Radius to peak & {Total gas mass} & Total GDR \\
$\mathrm{}$ & ($\mathrm{K}$) & $\mathrm{}$ & {($\mathrm{M_{\odot}}$)} & {($\mathrm{M_{\odot}}$)} & $\mathrm{}$ & ($\mathrm{{}^{\circ}}$) & ($\mathrm{{}^{\circ}}$) & ($\mathrm{kpc}$) & {($\mathrm{M_{\odot}}$)} & $\mathrm{}$ \\
\hline
164 & 13.6 & 2.4 & 1.3E+06 & 5.2E+04 & 24.0 & 11.1 & 41.6 & 6.0 & 3.8E+05 & 31.3 \\
313 & 17.5 & 2.3 & 1.6E+05 & 6.4E+03 & 25.6 & 11.2 & 41.9 & 10.8 & 1.9E+05 & 55.4 \\
159 & 15.6 & 2.2 & 7.0E+05 & 3.2E+04 & 22.1 & 11.1 & 41.6 & 6.7 & 6.5E+05 & 42.6 \\
323 & 17.8 & 2.2 & 3.5E+05 & 1.2E+04 & 29.2 & 11.1 & 41.9 & 11.2 & 4.7E+05 & 68.4 \\
150 & 15.5 & 2.2 & 2.0E+05 & 7.3E+03 & 27.3 & 11.1 & 41.6 & 6.0 & 9.4E+04 & 40.1 \\
146 & 15.1 & 2.1 & 5.5E+05 & 3.5E+04 & 15.6 & 11.4 & 41.9 & 11.9 & 1.4E+06 & 54.1 \\
241 & 13.9 & 2.3 & 2.1E+05 & 1.0E+04 & 20.2 & 11.3 & 41.9 & 10.3 & 2.1E+05 & 40.7 \\
321 & 17.6 & 2.2 & 2.4E+05 & 9.3E+03 & 25.9 & 11.1 & 41.9 & 11.5 & 3.5E+05 & 64.0 \\
253 & 13.8 & 2.4 & 1.4E+05 & 7.2E+03 & 19.6 & 11.3 & 41.9 & 10.4 & 1.3E+05 & 37.4 \\
310 & 15.3 & 2.1 & 4.6E+05 & 1.4E+04 & 32.2 & 11.2 & 41.9 & 11.7 & 4.4E+05 & 62.8 \\
\end{tabular}
\end{table}

\end{landscape}
\end{appendix}


\end{document}